%% file: manuscript.tex
\documentclass[a4paper]{packages/paper_cw} 
\input{packages/packages.tex}

\makenomenclature
\input{packages/definitions.tex}

\input{packages/statFEM_definitions.tex}
\begin{document}
\title{Mechanical State Estimation with a Polynomial-Chaos-Based Statistical Finite Element Method
}
\author{Vahab Narouie$^{\star,1}$, Henning Wessels$^{1}$, Fehmi Cirak$^{3}$, Ulrich R{\"o}mer$^{2}$}
\institute{(1) Institute of Applied Mechanics, Division Data-driven Modeling of Mechanical Systems, Technische Universität Braunschweig, Pockelsstr. 3, 38106 Braunschweig, Germany \\
	\noindent(2) Institute for Acoustics and Dynamics, Technische Universität Braunschweig, Langer Kamp 19, 38106 Braunschweig, Germany \\
	\noindent(3) Department of Engineering, University of Cambridge, Trumpington Street, Cambridge CB2 1PZ, United Kingdom \\
	$\star$ \email{\href{mailto:v.narouie@tu-braunschweig.de}{v.narouie@tu-braunschweig.de}} \\
}
\maketitle
\thispagestyle{empty}

\vspace{-2cm}
\graphicalabstract{
	\vspace{-1cm}
	\begin{figure}[H]
		\centering
		\includegraphics{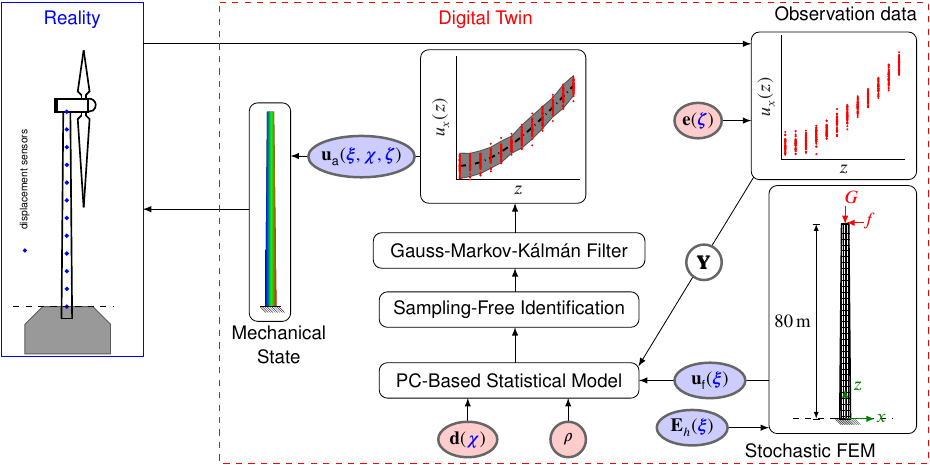}
	\end{figure}
}
\abstract
{
The Statistical Finite Element Method (statFEM) offers a Bayesian framework for integrating computational models with observational data, thus providing improved predictions for structural health monitoring and digital twinning. This paper presents a sampling-free statFEM tailored for non-conjugate, non-Gaussian prior probability densities. We assume that constitutive parameters, modeled as weakly stationary random fields, are the primary source of uncertainty and approximate them using the Karhunen-Lo{\`e}ve (KL) expansion. The resulting stochastic solution field, i.e., the displacement field, is a non-stationary, non-Gaussian random field, which we approximate via the Polynomial Chaos (PC) expansion. The PC coefficients are determined through projection using Smolyak sparse grids.
Additionally, we model the measurement noise as a stationary Gaussian random field and the model misspecification as a mean-free, non-stationary Gaussian random field, which is also approximated using the KL expansion and where the coefficients are treated as hyperparameters. The PC coefficients of the stochastic posterior displacement field are computed using the Gauss-Markov-K{\'a}lm{\'a}n filter, while the hyperparameters are determined by maximizing the marginal likelihood. We demonstrate the efficiency and convergence of the proposed method through one- and two-dimensional elastostatic problems.
}

\section{Introduction}\label{sec:Introduction}
As modern automation and information technologies continue to advance, the integration of sensors \cite{he2022integrated}, communication systems \cite{lu2015information}, and computer technologies with traditional structural engineering has become a key area of research. Monitoring of critical infrastructures such as bridges \cite{gharehbaghi2022critical,azimi2020data}, dams \cite{bukenya2014health}, and wind turbines \cite{prasad2023robust} is pressingly needed for early warnings and detection of possible defects or damages. Therefore, using a network of sensors, such as strain gauges or \textit{Fiber Bragg Gratings} (FBGs) \cite{sierra2016damage} is a common practice to monitor dynamic structural responses and is suitable for real-time \textit{Structural Health Monitoring} (SHM). To facilitate real-time monitoring, mechanical state estimation, and predictive analysis, \textit{Digital Twin} (DT) technologies have attracted considerable interest due to their capability to provide virtual replicas of physical systems \cite{torzoni2024digital}. A key component of Digital Twins in mechanical systems is the integration of state estimation techniques, such as the K{\'a}lm{\'a}n Filter, which addresses the challenge of uncertainty in physical systems and measurement noise \cite{feng2023model}. Moreover, Digital Twins facilitate continuous data exchange between physical and virtual models, which enhances real-time decision-making and optimizing system performance \cite{thelen2022comprehensive}.

Modeling physical systems using \textit{Partial Differential Equations} (PDEs) is fundamental for describing phenomena like fluid dynamics, heat transfer, and structural mechanics. Solving these complex equations needs numerical methods like the \textit{Finite Element Method} (FEM) \cite{eduardo1968theoretical}. The chosen material model and the associated, only partially known, parameters represent a major source of uncertainty in the \textit{Finite Element} (FE) model. Therefore, over the past two decades, researchers have focused more and more on developing different stochastic methods which can be integrated with FEM, commonly referred to as \textit{Stochastic Finite Element Method} (SFEM) \cite{stefanou2009stochastic}. Among these methods, \textit{Monte Carlo} (MC) simulation \cite{papadrakakis1996robust} and its enhanced versions \cite{graham2011quasi,badia2021embedded} have become more popular, mostly because they are simple to implement and can use the existing deterministic FEM solvers. This non-intrusive nature makes MC techniques especially valuable for addressing stochastic problems in FEM without needing much modification to the underlying solvers. However, getting highly accurate stochastic solutions with MC simulations requires many deterministic runs, making it computationally infeasible, especially for large or nonlinear stochastic problems. To address this limitation, different methods like \textit{Polynomial Chaos} (PC) expansion \cite{sudret2000stochastic,keese2003review,ghanem2003stochastic}, \textit{generalized Polynomial Chaos} (gPC) expansion \cite{xiu2002wiener,xiu2002modeling}, \textit{Gaussian processes} (GPs) \cite{mackay1998introduction,seeger2004gaussian} and manifold methods \cite{soize2021probabilistic} were developed to approximate the stochastic solution more efficiently. A widely used approach is the PC expansion, based on the theory of homogeneous chaos proposed by Wiener \cite{wiener1938homogeneous}. PC expansions capture the stochastic behavior of the physical system efficiently by approximating the random variables as a finite series of orthogonal polynomials, adapted to the input probability distribution. Several approaches have been developed to compute the coefficients of the PC expansion. These approaches can be grouped into four main categories: (1) Galerkin Polynomial Chaos \cite{babuska2004galerkin,matthies2005galerkin}, which determines the PC coefficients using a Galerkin projection in the probability space of the stochastic boundary value problem; (2) Stochastic Regression \cite{Berveiller2005,berveiller2006stochastic}, which determines the PC coefficients using least-squares minimization over a set of input samples; (3) Pseudo-Spectral Projection \cite{xiu2002wiener,reagana2003uncertainty}, which computes the PC coefficients using quadrature points; and (4) Stochastic Collocation \cite{xiu2005high,babuvska2007stochastic}, where the PC expansion is fitted by evaluating solutions at a set of collocation points chosen based on the distribution of the random inputs. The application of SFEM has been studied in many disciplines. For example, in \cite{sudret2002comparison}, the MC simulation and PC expansion are compared \Rev{for evaluating the probability of failure in a linear elastic problem}. In \cite{feng2021performance}, several SFEM methods were compared in linear and nonlinear mechanics, including plasticity and damage.

Although PC expansion effectively captures the stochastic solutions of physical systems, it has difficulties working with high-dimensional input spaces. This problem, called the curse of dimensionality, happens because the number of polynomial basis terms and unknown PC coefficients grows exponentially when the number of input random variables increases. To this end, methods like sparse quadrature \cite{smolyak1963quadrature,gerstner1998numerical,novak1999simple} were developed to reduce the computational cost of evaluating multidimensional integrals. Moreover, techniques such as \textit{least angle regression} (LAR) \cite{blatman2010adaptive,blatman2011adaptive} and compressive sensing \cite{doostan2011non,hampton2015compressive} provide sparse representations of the stochastic solution by choosing only the most important polynomial bases.  The theory and practical basis for using Smolyak sparse grids to estimate integrals of high-dimension is proposed in \cite{kaarnioja2013smolyak}. This approach uses tensor products of successive differences in univariate rules, which reduce the number of integration points compared with the traditional full product methods.

The task of identifying uncertain parameters using the observational data has been much researched for different materials, such as elastic \cite{Mahnken2017,hartmann2018identifiability}, hyperelastic \cite{hossain2013more}, and elastoplastic \cite{mahnken1997parameter,cooreman2007elasto,mathieu2015estimation} material models. However, the current modeling methods face difficulties in combining the measurement data with uncertainties and the predictions from the FE models, which are inherently misspecified. At the same time, the increasing number of sensor data has led to the development of new data-driven methods \cite{montans2019data}, which are changing the traditional paradigms in mechanics and material science. For example, in \cite{le2015computational}, the multiscale analysis of heterogeneous materials is studied using neural networks. Still, even the most advanced data-driven models need much real training data to create accurate predictions. Because of this, there is a growing need to combine FE models with data-driven methods to offer a more complete virtual representation of real physical systems.

The \textit{statistical finite element method} (statFEM) framework \cite{girolami2021statistical,akyildiz2022statistical,narouie2023inferring,koh2023stochastic} was introduced to address these challenges. It allows predictions about a system's true behavior even when sensor data are limited, and the FE model is misspecified. Similar to other Bayesian approaches, statFEM represents uncertainties \textemdash either due to noise in the collected data or the selection of the FE model \textemdash as random variables. The Bayes' rule gives a consistent way to infer the posterior distribution of these random variables, starting from an assumed prior distribution and considering the likelihood of the observed data \cite{gelman2013bayesian,stuart2010inverse}. In the first works on statFEM \cite{girolami2021statistical,febrianto2022digital}, the observed data was decomposed into three random components: an FE part, a model misspecification part, and noise from the measurements. The noise from the measurements and model misspecification were modeled as GPs. The misspecification component had three unknown hyperparameters that must be learned from the observed data. This learning process was done using a Bayesian approach by maximizing the marginal likelihood.

In the original formulation, the prior density resulting from uncertainties on the right-hand side of the PDE is modeled with a GP. The prior probability density for the FEM part is computed by applying a conventional stochastic FEM. However, one of the main difficulties in Bayesian inference for PDE-based problems with uncertainties is that the posterior density is not always analytically tractable. This gets more difficult when the priors are non-conjugate, meaning that the posterior distribution is not of the same family as the prior. In these situations, we need to use other approximation techniques, like the \textit{Markov Chain Monte Carlo} (MCMC) \cite{gilks1995markov,madras2002lectures,gamerman2006markov,rosic2013parameter}, Laplace approximation \cite{mackay1998choice}, or \textit{Variational Bayes} (VB) \cite{jordan1999introduction,blei2017variational,povala2022variational,archbold2024variational} to estimate or approximate the posterior distribution. This issue often occurs in Bayesian updating with stochastic FEM because uncertainties can originate not only from the right-hand side of the PDE, like external forces, but also from uncertain material parameters in the selected material model. Even when the uncertainties in material parameters are Gaussian, the associated prior density of the solution is usually not Gaussian.

As mentioned before, a common method for approximating the posterior distribution is MCMC, where the equilibrium distribution of the Markov chain represents the target posterior distribution by a random walk. However, MCMC often has slow convergence and needs a burn-in period to ensure the chain reaches the equilibrium. Also, MCMC can be very expensive to compute, especially for large-scale problems. In many engineering applications, observational data are often time- and space-dependent, providing different information at different time steps or locations. For such non-stationary problems, Bayesian filtering methods are often more suitable, as they estimate the \textit{Conditional Expectation} (CE) \cite{vondrejc2019accurate} of the system state recursively, given the data. The most well-known method in this category is the \textit{K{\'a}lm{\'a}n filter} \cite{kaipio2006statistical,chui2017kalman}, which is highly efficient for linear systems with Gaussian noise.

\Rev{For nonlinear and time-dependent PDEs, the filtering problem becomes significantly more challenging due to the nonlinearity of the governing equations and the presence of non-Gaussian stochastic distributions, which are difficult to approximate accurately. A promising alternative in such cases is the \textit{ensemble K{\'a}lm{\'a}n filter} (EnKF), which employs a sampling-based approximation to handle non-Gaussianity and dynamically update the stochastic model by sequentially assimilating observational data. This approach has been successfully adopted in the context of statFEM, as demonstrated in \cite{duffin2021statistical}. The unscented K{\'a}lm{\'a}n filter is also used in \cite{diaz2023new} to assimilate noisy sequential measurement data and recursively recover material state and model parameter. To further enhance the filtering framework,} the \textit{Gauss-Markov-K{\'a}lm{\'a}n filter} (GMKF) has been proposed as an \Rev{extension of the classical K{\'a}lm{\'a}n filter} in \cite{matthies2016inverse,rosic2013parameter} by representing all random variables in terms of functional approximations in \cite{ernst2014bayesian,matthies2016bayesian}. Another way to approximate the CE is by the PC-based GMKF, which is a sampling-free approach for the filtering problems \cite{rosic2012sampling,pajonk2013sampling}. However, the initial GMKF formulation struggles to capture the posterior distribution outside the linear, Gaussian setting \cite{ernst2015analysis}, prompting refined versions in recent years, such as those presented in \cite{vondrejc2019accurate}. \Rev{More recently, machine learning-based approaches have gained attention as an alternative to classical filtering methods. A \textit{machine learning-based conditional mean filter} has been proposed in \cite{hoang2023machine}, aiming to improve the efficiency and flexibility of Bayesian filtering for complex systems.}

The primary purpose of this paper is to provide a GMKF-based approach to the statFEM \cite{girolami2021statistical}, where prior and posterior are fully described via the PC coefficients of the displacement. A PC approach to statFEM has already been put forth in \cite{narouie2023inferring}; however, the PC method there was restricted to the computation of the prior\footnote{\Rev{To promote reproducibility, the source code associated with that work is now available in \cite{narouie2025statfem_recon}.}}. The GMKF directly results in a posterior update of the PC coefficients and provides a natural extension to handle non-conjugate priors. We additionally propose an extension for the model-reality mismatch that may provide an indication of structural damage in a monitoring context. More precisely, the model's misspecification is described by a KL expansion, where each mode has unknown parameters, which makes it a non-stationary random field. These unknown parameters are treated as hyperparameters and identified based on observational data by maximizing the surrogate marginal likelihood. The inference of hyperparameters is done numerically using Smolyak sparse grids to calculate the marginal likelihood efficiently. The resulting GMKF-statFEM method is sampling-free, reducing the need for expensive online FEM calculations.

The remainder of this manuscript is structured as follows: In \autoref{sec:Prior_of_Statistical_Finite_Element_Method}, the stochastic prior model is presented. We consider the balance of linear momentum and uncertain material parameters. The latter are modeled as a weakly stationary random field. The stochastic displacement field is approximated by a PC expansion. In \autoref{sec:Polynomial-Chaos-based_StatFEM}, the statistical generating model in its PC expansion is introduced. The updating scheme, involving the GMKF and the identification of hyperparameters, is explained in detail. We consider 1D and 2D mechanical test cases in \autoref{sec:Numerical_Examples} as numerical results. The paper concludes in \autoref{sec:Conclusion_Outlook}.

We introduce the following conventions for mathematical symbols. Scalars and scalar-valued functions are written in italic font, e.g., $\sclA$, $\scla$, $\sclB$, $\sclb$, $\varrho$, $\Psi$, and $\psi$, and include both uppercase and lowercase Roman and Greek letters. Column vectors and vector-valued functions are represented in boldface Roman fonts, such as $\vecA$, $\veca$, $\vecB$, $\vecb$, $\vecvarrho$, $\vecPsi$, $\vecpsi$, and $\vecsigma$. Matrices and matrix-valued functions are denoted using double boldface Roman font, e.g., $\matA$, $\mata$, $\matB$, $\matb$, $\matvarrho$, $\matPsi$, $\matpsi$, and $\matsigma$. Second-order tensors and tensor-valued functions follow the same convention as matrices, using double boldface Roman font. Sets are represented using blackboard bold fonts, such as $\bbR$, $\bbU$, and $\bbY$; otherwise, they are explicitly stated in the text. $\sigma$-algebras are denoted by script typestyle for uppercase letters, e.g., $\mathscr{F}$, $\mathscr{B}$.

\section{Prior of the Statistical Finite Element Method}\label{sec:Prior_of_Statistical_Finite_Element_Method}
Statistical finite element analysis is comprised of three steps: the computation of the prior, the identification of hyperparameters, and the computation of the posterior or inference. This section concentrates on the statFEM prior, which is essentially a \textit{stochastic finite element method} (SFEM) solution. The definitions of random variables and random fields are presented in \autoref{subsec:Definition_Of_Random_Variables_And_Random_Fields}. \autoref{subsec:Stochastic_Boundary_Value_Problem_and_Karhunen-Loeve_Expansion} presents the stochastic mechanical boundary value problem and the Karhunen-Lo\`eve expansion of the random field. \autoref{subsec:Polynomial_Chaos_Expansion_of_Random_Fields} elucidates the polynomial expansion of the simulation's output response.

\subsection{Definition of Random Variables and Random Fields}\label{subsec:Definition_Of_Random_Variables_And_Random_Fields}

Consider a \textit{probability space} denoted as $(\saSp, \sigAl, \prb)$, where $\saSp$ represents the \textit{sample space}, $\sigAl$ a \textit{$\sigma$-algebra} and $\prb:\sigAl \rightarrow [0,1]$ a scalar-valued \textit{probability measure}. For a parameter set, denoted as $\paraSpace \subset \bbR^n$, we consider the measurable space $(\paraSpace, \, \mathscr{B}_{\paraSpace})$ with the Borel \textit{$\sigma$-algebra} $\mathscr{B}_{\paraSpace}$. A \textit{random variable} (RV) is defined as a measurable mapping $H: (\saSp, \sigAl) \rightarrow (\bbR, \mathscr{B}_{\bbR})$, whereas an $M$-dimensional random vector is defined as $\mathbf{H} : (\saSp, \sigAl) \rightarrow (\bbR^M, \mathscr{B}_{\bbR^M})$. With this notation at hand, we can instead consider the probability space $\left(\bbR^M, \mathscr{B}_{\bbR^M}, f_{\mathbf{H}} d \mathbf{H} \right)$, where $f_{\mathbf{H}}$ denotes the associated (often $M$-dimensional Gaussian) probability density function.

The function space $L^p(\saSp, \prb), \,  p\geq 1,$ is the collection of random variables with a finite norm $\norm{H}_{L^p}^p = \int_{\saSp} \norm{H(\outcomeSaSp)}^p \ \prb(d \theta)$ where $\theta \in \saSp$ and the Hilbert space $L^2(\saSp,\prb)$ contains the so-called second order RVs. In the same way, we can define $L^p(\bbR^{M},f_{\mathbf{H}} d \mathbf{H})$, whereas $L^p(\bbR^{M},f_{\mathbf{H}} d \mathbf{H};\bbR^{N})$ denotes the associated space of random variables mapping from $(\bbR^M,\mathscr{B}_{\bbR^M})$ into $(\bbR^N,\mathscr{B}_{\bbR^N})$ that have a finite $L^p$-norm. The operator $\expectOper : L^1(\saSp, \prb) \rightarrow \bbR$ is the mathematical expectation, denoted as $\expectOper[H] =  \mu_H = \int_{\saSp} H(\outcomeSaSp) \prb(d \outcomeSaSp)$.

A \textit{random field} is defined as a mapping $H: \physicalDomain \times \saSp \rightarrow \bbR$, where, in our case, $\physicalDomain$ represents a bounded physical domain, $\physicalDomain \subset \bbR^{\DimensionDomain}$ with $\DimensionDomain \in \{1,2,3\}$. A random field can be viewed as an (infinite) ensemble of RVs and the joint distribution of any finite number of these RVs, i.e., $ \{H(\spatialPoint_1, \outcomeSaSp), \dots, H(\spatialPoint_n, \outcomeSaSp) \suchAs H(\spatialPoint_i, \cdot) \in L^p(\saSp, \prb)  \}$, is known as the $\orderFiniteDim$-th order finite-dimensional distribution of random fields \cite{grigoriu2012stochastic, uribe2020bayesian}.

For Gaussian random fields, the $\orderFiniteDim$-th order finite-dimensional distribution is a multivariate Gaussian for each $n$. Gaussian random fields are fully described by their first and second-order moments. This encompasses the (continuous) mean function $\meanFunc{H}: \physicalDomain \rightarrow  \bbR$ where $\meanFunc[H]{\spatialPoints} = \expectOper[H(\spatialPoints, \cdot)] = \int_{\saSp} H(\spatialPoints, \outcomeSaSp) \prb(d \outcomeSaSp)$ and covariance function $\covFunc{H} : \physicalDomain^2 \rightarrow \bbR$, defined as
\begin{equation}
	\covFunc[H]{\spatialPoints} = \expectOper[\big( H(\spatialPoints, \cdot)- \meanFunc[H]{\spatialPoints} \big) \big(H(\spatialPoints', \cdot)- \meanFunc[H]{\spatialPoints'} \big)] = \stdFunc[H]{\spatialPoints} \stdFunc[H]{\spatialPoints'} \corFunc[H]{\spatialPoints}.
\end{equation}
In this context, $\stdFunc{H} : \physicalDomain \rightarrow \bbR_0^{+}$ represents the standard deviation function, and $\corFunc{H}{}: \physicalDomain^2 \rightarrow [-1, 1]$ denotes the correlation function of the random field with $\spatialPoints, \spatialPoints' \in \physicalDomain$. In this contribution, the covariance function $\covFunc[H]{\spatialPoints}$ is also called covariance kernel or kernel function interchangeably.

A random field is defined as strictly stationary when its corresponding finite-dimensional distributions remain unchanged under any spatial shifts, represented by  $\vecd = \spatialPoints - \spatialPoints'$. If the mean function $\meanFunc{H}(\spatialPoints)$ remains constant across space, but the covariance function depends only on the shift, meaning $\covFunc[H]{\spatialPoints} = \covFuncD{H}{\vecd}$, the field is deemed weakly stationary, see \cite{uribe2020bayesian,lindgren2012stationary, vanmarcke2010random}. In contrast, a non-stationary random field has a spatially varying mean function or a covariance function influenced by both the shift and the values of $\stdFunc[H]{\spatialPoints}$ and $\stdFunc[H]{\spatialPoints'}$.

\subsection{Stochastic Boundary Value Problem and Karhunen-Lo\`eve Expansion}\label{subsec:Stochastic_Boundary_Value_Problem_and_Karhunen-Loeve_Expansion}
We assume that material parameter $\kappa$ depends on a random vector $\setYoungRV = [\youngRV{1}, \dots, \youngRV{i}, \dots, \youngRV{\nKlRvOfYoung}]^\top$ with independent standard Gaussian RVs $\youngRV{i}: (\saSp, \sigAl) \rightarrow (\bbR,\mathscr{B}_{\bbR})$. The static \textit{stochastic boundary value problem} (SBVP) for deformations is then defined as seeking a stochastic function $\stochDisp(\spatialPoint,\setYoungRVs)$, such that
\begin{equation}
	\begin{cases}
		\nabla \cdot \firstPK \left(\spatialPoints,\kappa \left( \spatialPoints, \setYoungRVs \right) \right) + \vecb(\spatialPoints) = \vec0, & \text{in} \,\, \physicalDomain  \\
		\stochDisp(\spatialPoints,\setYoungRVs) = \preStochDisp(\spatialPoints),                                                               & \text{on} \,\, \preDispOnBound  \\
		\firstPK(\spatialPoints,\kappa( \spatialPoints, \setYoungRVs)) \cdot \unitNormal = \preTranction(\spatialPoints),                      & \text{on} \,\, \preTracOnBound,
	\end{cases}
	\label{eq:SBVP}
\end{equation}
holds almost everywhere. We will consider a finite element discretization together with a projection method in the stochastic space. Hence, \eqref{eq:SBVP} needs to be transformed into a variational formulation over the combined physical and stochastic space. Since the procedure is standard, details are omitted here. Here, $\firstPK$ denotes the \textit{first Piola-Kirchhoff} ($1$.PK) stress tensor,  $\vecb$ the body force, $\preStochDisp$ and $\preTranction$ are the prescribed displacement and traction on Dirichlet and Neumann boundaries $\preDispOnBound$ and $\preTracOnBound$, respectively, such that $\partial\physicalDomain = \preDispOnBound \cup \preTracOnBound$ and $\partial\physicalDomain_{\vecu} \cap \preTracOnBound =\vecEmpty$. The unit vector normal to boundary $\preTracOnBound$ is denoted by $\unitNormal$. Note that $\spatialPoints$ denotes the coordinates within the domain $\physicalDomain$ in the reference configuration. In this formulation, $\kappa$ is a stochastic function of random vector $\setYoungRVs$ representing the random variability in the material properties across the domain $\physicalDomain$, where we use the notation $\setYoungRVs$ instead of $\setYoungRV$ for brevity.

In this paper, the uncertain source of the SBVP is Young's modulus, modeled as a weakly stationary random field. To ensure positive realizations, we define $\youngRFs = \exp{\gaussRFs}$, where $\gaussRFs$ is a weakly stationary Gaussian random field. Using the \textit{Karhunen-Lo\`eve} (KL) expansion, $\gaussRFs$ is approximated as:
\begin{equation}
	\gaussRFs =  \meanFunc[\kappa]{\spatialPoints} + \stdOf{\kappa} \sum_{\scli=1}^{\infty} \sqrt{\eigenValue{i}} \, \eigenFunction[i]{\spatialPoints}\, \youngRVs{i} \approx \meanFunc[\kappa]{\spatialPoints} + \stdOf{\kappa} \sum_{\scli=1}^{\nKlRvOfYoung} \sqrt{\eigenValue{i}} \, \eigenFunction[i]{\spatialPoints} \, \youngRVs{i},
	\label{eq:kl_gaussian_random_variable}
\end{equation}
see \cite{karhunen1947ueber,loeve1948functions}. Here, $\eigenValue{i} \in [0,\infty)$, and $\eigenValue{1} \ge \eigenValue{2} \ge \cdots \ge 0$ are the non-increasing eigenvalues, whereas $\eigenFunction{\scli} : \physicalDomain \rightarrow \bbR$ are the eigenfunctions of the correlation function $\corFunc[\kappa]{\spatialPoints}$. Further details on the KL expansion can be found in \autoref{Appendix:Karhunen-Loeve_Expansion_of_Random_Fields}.

In a finite element computation, the domain $\physicalDomain$ is discretized into subdomains, referred to as elements, each with a centroid at position vector $\vecX_i$ where $i = 1, \dots, n_{\text{elm}}$, and $n_{\text{elm}} \in \mathbb{N}^+$ represents the number of elements. Then, the continuous random field $\gaussRFs$ is discretized to obtain the random vector $\gaussRVECs = \left[ \gaussRVARs{1}, \dots, \gaussRVARs{i}, \dots, \gaussRVARs{n_{\text{elm}}} \right]^\top$, and subsequently, the Gaussian and log-normal random vectors $\gaussRVECs$ and $\youngRVECs = \exp{\gaussRVECs}$, respectively. Moreover, the stochastic function $\stochDisp(\spatialPoint, \youngOutcomeS)$, when evaluated at a specific set of nodes $\{\vecX_1, \dots, \vecX_{\nnode}\}$, forms a random vector with values in $\mathbb{R}^{\ngdof}$, and the FE solution is denoted by $\stochDisp_{h}(\setYoungRVs) \in \mathbb{R}^{\ngdof}$, with $\ngdof = \nnode \cdot d$ indicating the number of global degrees of freedom. For clarity, the random vector $\stochDisp_h(\setYoungRVs)$ is represented in terms of its components as
$\stochDisp_h(\setYoungRVs) = \big( \stochDisp_h^1(\setYoungRVs)^\top, \dots , \stochDisp_h^d(\setYoungRVs)^\top \big)^\top$, where $\stochDisp_h^v(\setYoungRVs)$ are the component functions along each basis direction.

A significant limitation of the KL expansion is the necessity to compute the eigenpairs $\eigenPairs{i}$ from integral eigenvalue problems as specified in \eqref{eq:secondKindFredholm}, which are often difficult to solve analytically, except for certain correlation functions and simple domains. Consequently, for different geometries and kernels, there is a requirement for numerical methods to estimate the eigenpairs of the equation presented in \eqref{eq:secondKindFredholm}, which is discussed thoroughly in \cite{ghanem2003stochastic}.

\subsection{Polynomial Chaos Expansion of Random Fields}\label{subsec:Polynomial_Chaos_Expansion_of_Random_Fields}
The KL expansion is suitable when the eigenpairs $\eigenPairs{i}$ of the correlation function of a random field are known. However, in dealing with a stochastic response $\stochDisp(\spatialPoint,\youngOutcomeS)$, the mean $\meanFunc[\stochDisp]{\spatialPoints}$ and covariance function $\covFunc[\stochDisp]{\spatialPoints}$ are not initially known, thus making it unsuitable for a KL approximation. To address this limitation, Ghanem and Spanos \cite{ghanem2003stochastic} employed the concept of homogeneous chaos and developed the spectral Stochastic Finite Element Method (SFEM).

The vector $\stochDisp_{h}^v(\setYoungRVs)$ for $v = 1, \dots, \DimensionDomain$ can be expanded using the PC expansion, specifically employing probabilistic-type Hermite polynomials $\pcMultiHermite{}{\bullet}$ as described in \cite{ghanem2003stochastic}, as follows
\begin{equation}
	\stochDisp_{h}^v(\setYoungRVs) =  \sum_{\sclj=0}^{\infty} \pcCof{\stochDisp}{\sclj}^v \, \pcMultiHermite{\sclj}{\setYoungRVs} \approx \sum_{\sclj=0}^{\pcOrderDisp} \pcCof{\stochDisp}{\sclj}^v \, \pcMultiHermite{\sclj}{\setYoungRVs} = \dispPcCofMatrix  \dispPcHerVec, \quad \text{with} \quad v = 1, \dots, \DimensionDomain.
	\label{eq:Disp_PC_Expansion}
\end{equation}
Here, the vector $\setYoungRVs$ is referred to as the \textit{germ} or the \textit{stochastic degrees of freedom} of the uncertainty. Note that in \eqref{eq:Disp_PC_Expansion}, $\dispPcCofMatrix$ is a matrix containing the PC coefficients of stochastic displacement for $v = 1,\dots, \DimensionDomain$, and $\dispPcHerVec = \left[ \pcMultiHermite{1}{\setYoungRVs}, \dots, \pcMultiHermite{\sclj}{\setYoungRVs}, \dots ,\pcMultiHermite{\pcOrderDisp}{\setYoungRVs} \right]^\top$ is a vector, which contains all the multivariate Hermite polynomials of stochastic displacement. The multivariate Hermite polynomials $\mPsi_j$ are defined as a product of univariate Hermite polynomials $\psi_{\varrho_i^j}$ with a multi-index $(\varrho_1^j,\ldots,\varrho_{\nKlRvOfYoung}^j)$
\begin{equation}
	\mPsi_j(\setYoungRVs) = \prod_{i=1}^{\nKlRvOfYoung} \psi_{\varrho_i^j}(\youngRVs{i}).
	\label{eq:multivariatePC}
\end{equation}
To generate the sequence of multi-indices $\varrho_i^j$, we can either use the graded lexicographic ordering method \cite{xiu2010numerical} or the ball-box method  \cite{sudret2000stochastic}, as utilized in this contribution. Note that the total number of expansion terms $\pcOrderDisp+1$ in \eqref{eq:Disp_PC_Expansion} depends on $\nKlRvOfYoung$ and the chosen polynomial order $\sclp$, which can be determined as
\begin{equation}
	\pcOrderDisp = \frac{(\nKlRvOfYoung+\sclp)!}{\nKlRvOfYoung! \, \sclp!} -1,
	\label{eq:POrder}
\end{equation}
see e.g. \cite{ghanem2003stochastic, sudret2000stochastic}. The PC coefficients $\pcCof{\stochDisp^v}{\sclj}$ in \eqref{eq:Disp_PC_Expansion} can be obtained by the pseudo spectral projection method \cite{sullivan2015introduction} as
\begin{equation}
	\pcCof{\stochDisp}{\sclj}^v = \frac{\dobInner{\stochDisp_{h}^v (\setYoungRVs)}{\pcMultiHermite{\sclj}{\setYoungRVs}}}{\hermiteDobInner{\sclj}{\setYoungRVs}} \approx \frac{1}{\hermiteDobInner{\sclj}{\setYoungRVs}}\sum_{\scln = 1}^{\nGL}  \stochDisp_{h}^v(\setYoungRVs_n) \,  \pcMultiHermite{\sclj}{\setYoungRVs_n} \, \wGL , \quad \text{with} \quad v = 1, \dots, \DimensionDomain,
	\label{eq:PC_Coof_identify}
\end{equation}
where $\nGL$ is the number of Gauss-Legendre quadrature points and $\setYoungRVs_n$ and $\wGL$ are the integration points and weights. The term $\stochDisp_{h}^v(\setYoungRVs_n)$ is the displacement obtained from an FE analysis based on $\youngRVECs$ evaluated at the Gauss-Legendre quadrature points $\setYoungRVs_n$. In this contribution, we used sparse Smolyak grids with a Gaussian Hermite quadrature rule because of the high dimensionality of our problem caused by the KL expansion, see \cite{smolyak1963quadrature,kaarnioja2013smolyak}. By back-substituting the identified PC coefficients $\pcCof{\stochDisp}{\sclj}^v$ into \eqref{eq:Disp_PC_Expansion} and sampling from $\setYoungRVs$, we can generate samples from the stochastic displacement function $\stochDisp_{h}^v(\setYoungRVs)$. This can be interpreted as either prior or forecasted displacement $\stochDisp^v_{\text{f}}$ in the Bayesian framework, which will be further utilized in the subsequent section.

The mean vector and the covariance matrix of the displacements are then defined as
\begin{equation}
	\begin{split}
		& \meanVec{\stochDisp^v}  = \pcCof{\stochDisp}{0}^v, \quad \text{with} \quad v = 1, \dots, \DimensionDomain, \\
		& \covMat{\stochDisp^v} = \sum_{\sclj=1}^{\infty} \hermiteDobInner{\sclj}{\setYoungRVs}  \pcCof{\stochDisp}{\sclj}^v \, (\pcCof{\stochDisp^v}{\sclj})^\top \approx \sum_{\sclj=1}^{\pcOrderDisp} \hermiteDobInner{\sclj}{\setYoungRVs}  \pcCof{\stochDisp}{\sclj}^v  \, (\pcCof{\stochDisp^v}{\sclj})^\top.
	\end{split}
	\label{eq:Mean_and_Cov_Disp_PC_Expansion}
\end{equation}

\section{Polynomial-Chaos-Based StatFEM}\label{sec:Polynomial-Chaos-based_StatFEM}
This section introduces the proposed sampling-free statFEM algorithm. The foundation for this method is the statistical generating model discussed in \autoref{sec:Statistical_Generating_Model}. In \autoref{sec:Bayesian_Inference_via_Filtering}, the Bayesian inference based on filtering is explained, and the procedure of updating PC coefficients of the stochastic prior via a Gauss-Markov-K{\'a}lm{\'a}n filter is discussed in \autoref{sec:Polynomial-Chaos-Based_Filter}. Identifying hyperparameters is thoroughly detailed in \autoref{sec:identify_hyperparameters}.

\subsection{Statistical Generating Model}\label{sec:Statistical_Generating_Model}
A central component of the statFEM is the statistical generating model that relates observation data with forecasted displacements. Possible observed values can be modeled by applying an observation operator $\vecO: (\bbR^{\ngdof},\mathscr{B}_{\bbR^{\ngdof}}) \to (\bbR^{\ngsen},\mathscr{B}_{\bbR^{\ngsen}})$ to the prior displacements to obtain $\obsOperator{\stochDisp_{\text{f}}(\setYoungRV)}$. Here, $\ngsen = \nsen \cdot \DimensionDomain$, where $\nsen \in \bbN^+$ is the number of sensors. However, in reality, model errors are non-negligible, and observations are contaminated with measurement noise.

Thus, the statistical model assumes that the forecasted observation data are comprised of three distinct stochastic processes as
\begin{equation}
	\begin{split}
		\surrSingleObs^v\big(\setYoungRVs, \setMrmRVs, \setNoisRVs \big) & = \rhoCof^v \, \Hproj^v \, \stochDisp_{h}^v(\setYoungRVs) + \stochMrm^v(\setMrmRVs) + \stochNoise^v(\setNoisRVs) \\
		& = \trueDisp^v(\setYoungRVs, \setMrmRVs) + \stochNoise^v(\setNoisRVs),
		\label{eq:statistical_generating_model2}
	\end{split}
\end{equation}
written for each spatial dimension $v = 1, \dots, \DimensionDomain$. In \eqref{eq:statistical_generating_model2}, the observation operator is represented by dimension-specific observation matrices $\Hproj^v \in \bbR^{\nsen \times \nnode}$, projecting the forecasted displacement $\stochDisp^v_{\text{f}} = \stochDisp_{h}^v$ onto the sensor locations. The scaling factors $\rhoCof^v$ and $\stochMrm^v$ account for the model deficiencies, whereas $\stochNoise^v$ represents measurement noise. For simplicity, we use the notations $\setYoungRVs,\setMrmRVs,\setNoisRVs$ instead of $\setYoungRV,\setMrmRV,\setNoisRV$, respectively. The three (discrete) stochastic processes occurring in \eqref{eq:statistical_generating_model2} are: The stochastic displacements $\stochDisp_{h}^v(\setYoungRVs)$ modeled as a non-stationary non-Gaussian random field; the model-reality mismatch $\stochMrm^v(\setMrmRVs)$, described as a non-stationary Gaussian random field and parameterized by a different set of random variables $\setMrmRVs$; the noise $\stochNoise^v(\setNoisRVs)$, modeled as a stationary Gaussian field, parameterized by the random vector $\setNoisRVs$. This expression clarifies that the forecasted observation data $\surrSingleObs^v \in \bbR^{\nsen}$ is dependent on the random variables $\setYoungRVs$, $\setMrmRVs$, and $\setNoisRVs$. Note that $\trueDisp^v \in \bbR^{\nsen}$ represents the true displacement at the sensor locations in each respective dimension.

\begin{figure}[!htb]
	\centering
	\includegraphics{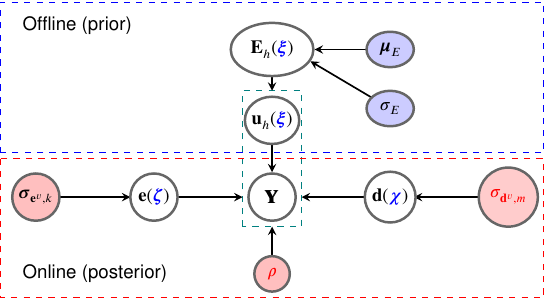}
	\caption{\textbf{PC-Based StatFEM Graphical Illustration:} The blue random variables $\setYoungRVs$ represent the stochastic Young's modulus and prior displacement, while $\setMrmRVs$ parameterize the model-reality mismatch, and $\setNoisRVs$ account for the noise. The red parameters $\rhoCof$ and $\mrmKlCof{\sclm}$ are unknown hyperparameters.}
	\label{fig:graphical_illustration}
\end{figure}

The goal is to update the stochastic displacement $\stochDisp^v_{\text{f}} = \stochDisp_{h}^v$ (e.g., its mean, standard deviation, etc.) in light of observation data $\SingleObs \in \bbR^{\ngsen}$. In practical applications, we can read $\nrep \in \bbN^+$ sets of observation data from the same sensors. Therefore, we collect the set of observations $\{\SingleObs\}_{r=1}^{\nrep}$ in a matrix $\multipleObs \in \bbR^{\ngsen \times \nrep}$. A graphical illustration of this model, adapted from \cite{girolami2021statistical}, can be found in \autoref{fig:graphical_illustration}.

\subsection{Bayesian Inference via Filtering}\label{sec:Bayesian_Inference_via_Filtering}
\begin{figure}[!htb]
	\centering
	\includegraphics{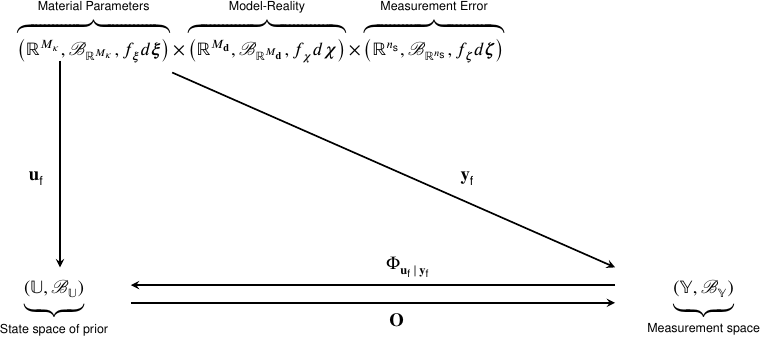}
	\caption{Diagram of Measurement spaces and their Mappings. This diagram is adapted from \cite{vondrejc2019accurate}.}
	\label{fig:Mapping_diagram}
\end{figure}

In a probabilistic setting, updating the stochastic displacement field is based on Bayesian inference, which relies on conditioning on observations. It is, in fact, a transition from the prior (or forecasted) probabilistic function of stochastic displacement $\stochDisp_{\text{f}}$ in \eqref{eq:Disp_PC_Expansion}  to a posterior (or assimilated) probabilistic function $\stochDisp_{\text{a}}$ using Bayes's theorem. We will temporarily omit the dimension-specific notation $v$ for this part to simplify the discussion. Assuming the prior stochastic displacement $\stochDisp_{\text{f}}$ and observation $\multipleObs$ have a joint probability density function (PDF), the posterior distribution of the stochastic displacement can be expressed in the Bayesian formalism as
\begin{equation}
	f_{\mathbf{u}_{\text{a}}}(\stochDisp) = f_{\stochDisp_{\text{f}} |  \, \multipleObs}(\stochDisp |  \, \multipleObs) = \frac{\PDF{\vecY | \mathbf{u}_{\text{f}}}{\multipleObs \, | \, \stochDisp} \, \PDF{\mathbf{u}_{\text{f}}}{\stochDisp}}{\PDF{\vecY}{\multipleObs}}, \ \quad \PDF{\vecY}{\multipleObs} >0, \ \stochDisp \in \bbU.
	\label{eq:Bayesian_formel}
\end{equation}
In this expression, $f_{\mathbf{u}_{\text{a}}}(\bullet) = f_{ \mathbf{u}_{\text{f}} \ |  \, \multipleObs}(\bullet |  \, \multipleObs)$ represents the assimilated or posterior probability density, while $f_{\mathbf{u}_{\text{f}}}$ denotes the forecasted or prior probability density of the stochastic displacement. The likelihood function $f_{\vecY | \mathbf{u}_{\text{f}}}(\mathbf{Y}| \bullet)$ evaluates how well the statistical model fits the observation data $\multipleObs$.  It is derived under the assumption that the observation data consists of independent and identically distributed (i.i.d.) samples, expressed as
\begin{equation}
	\PDF{\vecY | \mathbf{u}_{\text{f}}}{\multipleObs \, | \, \stochDisp} = \prod_{r=1}^{\nrep} \PDF{\vecy_r | \mathbf{u}_{\text{f}} }{\SingleObs \, | \, \stochDisp}.
	\label{eq:likelihood_from_single_observation}
\end{equation}
The marginal likelihood $\PDF{\vecY}{\multipleObs}$ is a normalizing constant that makes the posterior density integrate into unity and is defined as
\begin{equation}
	\PDF{\vecY}{\multipleObs} =  \int  \prod_{r=1}^{\nrep} \PDF{\vecy_r | \mathbf{u}_{\text{f}} }{\SingleObs \, | \, \stochDisp} \, \PDF{\mathbf{u}_{\text{f}}}{\stochDisp} \, \mathrm{d} \stochDisp.
	\label{eq:marginal_likelihood}
\end{equation}
The posterior density in \eqref{eq:Bayesian_formel} can be computed using sampling techniques such as \textit{Markov-chain Monte Carlo} (MCMC) \cite{gilks1995markov,gamerman2006markov}. However, for the most part, there is no need to compute the entire posterior density. It is often sufficient to derive estimates of the most significant statistical properties of the posterior PDF of stochastic displacement, such as the posterior mean function $\meanFunc[\stochDisp_{\text{a}}]{\spatialPoints}$ and posterior covariance function $\covFunc[\stochDisp_{\text{a}}]{\spatialPoints}$ of the stochastic displacement based on the conditional probability. To this end, \eqref{eq:Bayesian_formel}  can be reformulated using the updating (filtering) formula as
\begin{equation}
	\stochDisp_{\text{a}}(\setYoungRVs, \setMrmRVs, \setNoisRVs) = \Phi_{\stochDisp_{\text{f}} \, |\, \mathbf{y}_{\text{f}}} (\SingleObs) + \Big(\stochDisp_{\text{f}}(\setYoungRVs) - \Phi_{\stochDisp_{\text{f}} \, | \, \mathbf{y}_{\text{f}}}(\mathbf{y}_{\text{f}}(\setYoungRVs, \setMrmRVs, \setNoisRVs)) \Big),
	\label{eq:CE_filter}
\end{equation}
where
\begin{equation}
	\Phi_{\stochDisp_{\text{f}} \, |\, \mathbf{y}_{\text{f}}} = \argmin_{\textbf{w} \in L^2(\bbY,f_{\mathbf{y}_{\text{f}}} d \mathbf{y}_{\text{f}};\bbU)} \expectOper_{\setYoungRVs, \setMrmRVs, \setNoisRVs} \Big\| \mathbf{u}_{\text{f}} - \mathbf{w} \circ \mathbf{y}_{\text{f}} \Big\|^2
\end{equation}
is the conditional expectation $\Phi_{\stochDisp_{\text{f}} \, |\, \mathbf{y}_{\text{f}}} (\SingleObs) = \expectOper[\stochDisp_{\text{f}} \, | \, \SingleObs]$, see \cite{tarantola2005inverse,stuart2010inverse,vondrejc2019accurate}. In \eqref{eq:CE_filter}, the first term represents the information originating from observation data, and the second represents forecasted or prior information. The structure of these mappings is summarized in \autoref{fig:Mapping_diagram}.

According to the Doob-Dynkin Lemma 2.1.24 in \cite{bobrowski2005functional}, $\Phi_{\stochDisp_{\text{f}} \, |\, \mathbf{y}_{\text{f}}} (\mathbf{y})= \lebesgueFunction(\mathbf{y})$, where $\mathbf{y} \in \bbY$ and $\lebesgueFunction$ is a vector-valued Lebesgue measurable function. Rearranging \eqref{eq:CE_filter} and inserting the Lebesgue measurable functions yields
\begin{equation}
	\stochDisp_{\text{a}}(\setYoungRVs, \setMrmRVs, \setNoisRVs) = \stochDisp_{\text{f}}(\setYoungRVs) +  \lebesgueFunction(\SingleObs) - \lebesgueFunction \Big(\surrSingleObs(\setYoungRVs, \setMrmRVs, \setNoisRVs) \Big),
	\label{eq:CE_filter1}
\end{equation}
which is a generalized K{\'a}lm{\'a}n filter equation. For computational purposes, $\lebesgueFunction$ is approximated using linear (affine) functions, which enables the construction of the \textit{Gauss-Markov-K{\'a}lm{\'a}n filter} (GMKF) \cite{rosic2013parameter,matthies2016inverse}, as
\begin{equation}
	\lebesgueFunction \Big(\surrSingleObs(\setYoungRVs, \setMrmRVs, \setNoisRVs) \Big) := \kgain \,  \surrSingleObs(\setYoungRVs, \setMrmRVs, \setNoisRVs) + \vecb
	\label{eq:CE_filter2}
\end{equation}
where $\kgain$ and $\vecb$ are derived from the following mean-squared minimization problem
\begin{equation}
	\kgain^\star, \vecb^\star = \argmin_{\kgain , \vecb} \, \expectOper_{\setYoungRVs, \setMrmRVs, \setNoisRVs} \bigg[ \Big\| \stochDisp_{\text{f}}(\setYoungRVs) - \kgain \, \surrSingleObs(\setYoungRVs, \setMrmRVs, \setNoisRVs) - \vecb \Big\|^2 \bigg].
	\label{eq:CE_filter3}
\end{equation}
This results in the linear Gauss-Markov-K{\'a}lm{\'a}n filter equation
\begin{equation}
	\stochDisp_{\text{a}}(\setYoungRVs, \setMrmRVs, \setNoisRVs) =  \stochDisp_{\text{f}}(\setYoungRVs) + \kgain \, \Big(\SingleObs - \surrSingleObs(\setYoungRVs, \setMrmRVs, \setNoisRVs)\Big),
	\label{eq:GMKF1}
\end{equation}
where, $\big(\SingleObs - \surrSingleObs(\setYoungRVs, \setMrmRVs, \setNoisRVs)\big)$ is referred to as the innovation and $\kgain$ as the K{\'a}lm{\'a}n gain defined for $\nrep$ sets of observation data as
\begin{equation}
	\kgain  = \covMat{\stochDisp} \, \Hproj^\top  \bigg(\rhoCof  \, \Hproj  \,  \covMat{\stochDisp}  \, \Hproj^\top + \frac{\covMat{\stochMrm} + \covMat{\stochNoise}}{\nrep  \, \rhoCof} \bigg)^{-1}.
	\label{eq:kalman_gain}
\end{equation}

\subsection{Polynomial-Chaos-Based Filter}\label{sec:Polynomial-Chaos-Based_Filter}
To accelerate the performance of the GMKF, we substitute the forecasted stochastic displacement $\stochDisp_{\text{f}}$ with the PC expansion of the stochastic displacement \eqref{eq:Disp_PC_Expansion} into \eqref{eq:statistical_generating_model2}, which yields
\begin{equation}
	\surrSingleObs^v(\setYoungRVs, \setMrmRVs, \setNoisRVs) = \rhoCof^v \, \Hproj^v \,  \sum_{\sclj=0}^{\infty} \pcCof{\stochDisp}{\sclj}^v  \pcMultiHermite{\sclj}{\setYoungRVs} + \stochMrm^v(\setMrmRVs) + \stochNoise^v(\setNoisRVs), \quad \text{with} \quad v = 1, \cdots, \DimensionDomain.
	\label{eq:statistical_generating_model3}
\end{equation}
The model-reality mismatch $\stochMrm^v \in \bbR^{\nsen}$ is modeled as a mean-free non-stationary Gaussian random field on sensor locations, which we assume to be independent of $\stochDisp_{h}^v$ and we describe it with a KL expansion as
\begin{equation}
	\stochMrm^v(\setMrmRVs) =  \sum_{\sclm=1}^{\infty}  \mrmKlCof{\sclm} \, \sqrt{\eigenValue{m}} \, \eigenVec{m}\, \mrmRV{m} \approx  \sum_{\sclm=1}^{\nKlRvOfMrm}  \mrmKlCof{m} \, \sqrt{\eigenValue{m}} \, \eigenVec{m} \, \mrmRV{m} = \eigenMat  \diag(\setMrmKlCof{v}) \setMrmRVs = \mrmKlCofMatrix  \setMrmRVs.
	\label{eq:Mrm_KL_Expansion}
\end{equation}
Here, $\eigenValue{m} \in [0,\infty), \, \eigenValue{1} \ge \eigenValue{2} \ge \cdots \ge 0$ are the non-increasing eigenvalues and the matrix $\eigenMat \in \bbR^{\nsen \times \nKlRvOfMrm}$, with column vectors $\eigenVec{m}$, contains the eigenfunctions of a correlation function at the sensor coordinates. Moreover, the matrix $\mrmKlCofMatrix = \eigenMat  \diag(\setMrmKlCof{v})$ contains the product of (square-roots of) eigenvalues and eigenfunctions at the sensor locations. In this formulation, the vector $\setMrmKlCof{v} = [\mrmKlCof{1}, \dots, \mrmKlCof{\sclm}, \dots, \mrmKlCof{\nKlRvOfMrm}]^\top$ contains the standard deviations for each mode in the KL expansion, which makes the model-reality mismatch a mean-free non-stationary random field. Concurrently, the vector $\setMrmRVs = [\mrmRV{1}, \dots, \mrmRV{m}, \dots, \mrmRV{\nKlRvOfMrm}]^\top$ is filled with independent standard Gaussian RVs, capturing the uncertainty in the model-reality mismatch.  The covariance matrix is then obtained from
\begin{equation}
	\covMat{\stochMrm^v} =  \sum_{\sclm=1}^{\infty} \mrmKlCof{\sclm}^2 \, \eigenValue{m} \eigenVec{m} \,\eigenVec{m}^\top \approx \sum_{\sclm=1}^{\nKlRvOfMrm} \,\mrmKlCof{\sclm}^2 \,\eigenValue{m} \, \eigenVec{m} \, \eigenVec{m}^\top  = \mrmKlCofMatrix \, \mrmKlCofMatrix^\top.
	\label{eq:Cov_Mrm_KL_Expansion}
\end{equation}

Additionally, the noise $\stochNoise^v(\setNoisRVs) \in \bbR^{\nsen}$ is modeled as a stationary Gaussian random field at the sensor locations, defined by
\begin{equation}
	\stochNoise^v(\setNoisRVs) = \sum_{\sclk=1}^{\nsen} \noisPcCof{k} \, \noisRV{\sclk} = \noisPcCofMat \setNoisRVs,
	\label{eq:Nois_PC_Expansion}
\end{equation}
where $\setNoisRVs = \big[\noisRV{1}, \hdots, \noisRV{k}, \hdots, \noisRV{\nsen}\big]^\top$ consists of independent standard Gaussian RVs. The matrix $\noisPcCofMat \in \bbR^{\nsen \times \nsen}$ encapsulates the standard deviations for each sensor across every physical dimension and is structured as $\noisPcCofMat = \big[\noisPcCof{1}, \dots, \noisPcCof{k}, \dots, \noisPcCof{\nsen}\big]$. Each $\noisPcCof{k} \in \bbR^{\nsen}$ is detailed as
\begin{equation}
	\noisPcCof{k} = \big[\delta_{\sclk1}\sigma_{\stochNoise^v,1}, \hdots, \, \delta_{\sclk\nsen} \sigma_{\stochNoise^v,\nsen}\big]^\top,
	\label{eq:noisPcCofT}
\end{equation}
where $\delta_{k\nsen}$ is the Kronecker delta. The covariance matrix is then defined by
\begin{equation}
	\covMat{\stochNoise^v} = \sum_{\sclk=1}^{\nsen} \noisPcCof{k} \, \noisPcCofT{k} = \noisPcCofMat \, \noisPcCofMat^\top.
	\label{eq:Cov_Nois_PC_Expansion}
\end{equation}

All the PC basis functions are combined in one comprehensive basis vector to unify the different random vectors under a common basis. Initially, we define the extended PC basis vector $\exPcVec \in \bbR^{\exPcOrder}$ with $\exPcOrder = \pcOrderDisp + \nKlRvOfMrm + \nsen$, structured as
\begin{equation}
	\exPcVec = \left[\dispPcHerVec, \, \setMrmRVs, \, \setNoisRVs \right]^\top
\end{equation}
Following this definition, the new PC representations of $\stochDisp_{h}^v$, $\stochMrm^v$, $\stochNoise^v$, and the observation $\SingleObs$ on this extended basis are then formulated as
\begin{equation}
	\stochDisp_{h}^v ( \setYoungRVs, \setMrmRVs, \setNoisRVs )  = \sum_{\alpha=0}^{\exPcOrder} \exPcCof{\stochDisp}{\text{f},\alpha}^v  \, \exPcMultiHermite{\alpha}{ \setYoungRVs, \setMrmRVs, \setNoisRVs } = \exDispPcCofMatrix_{\text{f}}^v \, \exPcVec
	\label{eq:extended_pc_expansions_displacement}
\end{equation}

\begin{equation}
	\stochMrm^v    ( \setYoungRVs, \setMrmRVs, \setNoisRVs )  = \sum_{\alpha=0}^{\exPcOrder} \exPcCof{\stochMrm}{\alpha}^v   \, \exPcMultiHermite{\alpha}{\setYoungRVs, \setMrmRVs, \setNoisRVs} = \exMrmPcCofMatrix^v \, \exPcVec
	\label{eq:extended_pc_expansions_model_reality_mismatch}
\end{equation}

\begin{equation}
	\stochNoise^v  ( \setYoungRVs, \setMrmRVs, \setNoisRVs )  = \sum_{\alpha=0}^{\exPcOrder} \exPcCof{\stochNoise}{\alpha}^v \, \exPcMultiHermite{\alpha}{\setYoungRVs, \setMrmRVs, \setNoisRVs} = \exNoisPcCofMatrix^v \, \exPcVec
	\label{eq:extended_pc_expansions_noise}
\end{equation}

\begin{equation}
	\begin{split}
		\SingleObs^v     = \sum_{\alpha=0}^{\exPcOrder} \hat{\vecy}_{r,\alpha}^v  \, \exPcMultiHermite{\alpha}{\setYoungRVs, \setMrmRVs, \setNoisRVs} = \exObsPcCofMatrix^v  \, \exPcVec.
	\end{split}
	\label{eq:extended_pc_expansions_observation}
\end{equation}
The expressions \eqref{eq:extended_pc_expansions_displacement}-\eqref{eq:extended_pc_expansions_observation} are subject to the following considerations:
\begin{itemize}
	\item In \eqref{eq:extended_pc_expansions_displacement}, the matrix $\exDispPcCofMatrix_{\text{f}}^v \in \bbR^{\nnode \times \exPcOrder}$ stores the prior PC coefficients of displacement $\pcCof{\stochDisp}{\sclj}^v$. Specifically, the first $\pcOrderDisp$ columns of $\exDispPcCofMatrix_{\text{f}}^v$ contain the PC coefficients for the prior displacement $\stochDisp_h^v$, and the columns from $\pcOrderDisp + 1$ onwards are filled with zeros, as follows:
	      \begin{equation}
		      \exDispPcCofMatrix_{\text{f}}^v =
		      \begin{bmatrix}
			      \pcCof{\stochDisp}{0}^v & \dots & \pcCof{\stochDisp}{\pcOrderDisp}^v & \veczero & \dots & \veczero
		      \end{bmatrix}.
	      \end{equation}
	\item In $\exMrmPcCofMatrix^v \in \bbR^{\nsen \times \exPcOrder}$ \eqref{eq:extended_pc_expansions_model_reality_mismatch}, the columns from $\pcOrderDisp + 1$ to $\pcOrderDisp + \nKlRvOfMrm$ hold the columns of matrix $\mrmKlCofMatrix$ as specified in \eqref{eq:Mrm_KL_Expansion}, and the rest are zeroes, as follows:
	      \begin{equation}
		      \exMrmPcCofMatrix^v =
		      \begin{bmatrix}
			      \veczero & \dots & \veczero & \mrmKlCofVec{1} & \dots & \mrmKlCofVec{\nKlRvOfMrm} & \veczero & \dots & \veczero
		      \end{bmatrix}.
	      \end{equation}
	\item For $\exNoisPcCofMatrix^v \in \bbR^{\nsen \times \exPcOrder}$ in \eqref{eq:extended_pc_expansions_noise}, the columns from $\pcOrderDisp + \nKlRvOfMrm + 1$ to $\exPcOrder$ are filled with the columns from matrix $\noisPcCofMat$ as detailed in \eqref{eq:Nois_PC_Expansion}, with the remaining columns also set to zero, as follows:
	      \begin{equation}
		      \exNoisPcCofMatrix^v =
		      \begin{bmatrix}
			      \veczero & \dots & \veczero & \noisPcCof{1} & \dots & \noisPcCof{\nsen}
		      \end{bmatrix}.
	      \end{equation}
	\item Since the observation data $\SingleObs^v$ in \eqref{eq:extended_pc_expansions_observation} is deterministic, only the first column of $\exObsPcCofMatrix^v$ is allocated with $\SingleObs^v$, and the remaining columns set to zero, as follows:
	      \begin{equation}
		      \exObsPcCofMatrix^v =
		      \begin{bmatrix}
			      \SingleObs^v & \veczero & \dots & \veczero
		      \end{bmatrix}.
	      \end{equation}
\end{itemize}
By incorporating \eqref{eq:extended_pc_expansions_displacement}, \eqref{eq:extended_pc_expansions_model_reality_mismatch} and \eqref{eq:extended_pc_expansions_noise} into the statistical generating model \eqref{eq:statistical_generating_model2}, we obtain its formulation based on the extended basis  $\exPcMultiHermite{\alpha}{\setYoungRVs, \setMrmRVs, \setNoisRVs}$ as
\begin{equation}
	\surrSingleObs^v \big(\setYoungRVs, \setMrmRVs, \setNoisRVs \big) = \rhoCof^v \, \Hproj^v \, \sum_{\alpha=0}^{\exPcOrder} \exPcCof{ \stochDisp }{\text{f},\alpha}^v  \, \exPcMultiHermite{ \alpha }{ \setYoungRVs, \setMrmRVs, \setNoisRVs } + \sum_{\alpha=0}^{\exPcOrder} \exPcCof{\stochMrm}{\alpha}^v   \, \exPcMultiHermite{\alpha}{\setYoungRVs, \setMrmRVs, \setNoisRVs} + \sum_{\alpha=0}^{\exPcOrder} \exPcCof{\stochNoise}{\alpha}^v \, \exPcMultiHermite{\alpha}{\setYoungRVs, \setMrmRVs, \setNoisRVs}.
	\label{eq:extended_statistical_generating_model}
\end{equation}
Note that in \eqref{eq:GMKF1}, the assimilated displacement $\stochDisp_{\text{a}}$ is characterized with $\setYoungRVs$, $\setMrmRVs$, and $\setNoisRVs$ and can be expanded dimension-specific with extended basis as
\begin{equation}
	\stochDisp_{\text{a}}^v ( \setYoungRVs, \setMrmRVs, \setNoisRVs )  = \sum_{\alpha=0}^{\exPcOrder} \exPcCof{ \stochDisp }{\text{a},\alpha }^v  \, \exPcMultiHermite{ \alpha }{ \setYoungRVs, \setMrmRVs, \setNoisRVs } = \exDispPcCofMatrix_\text{a}^v \, \exPcVec.
	\label{eq:extended_pc_expansions_assimilated_displacement}
\end{equation}
Consequently, by replacing \eqref{eq:extended_pc_expansions_displacement}, \eqref{eq:extended_pc_expansions_observation},  \eqref{eq:extended_statistical_generating_model} and \eqref{eq:extended_pc_expansions_assimilated_displacement} in \eqref{eq:GMKF1}, we derive the PC representation of the GMKF  as follows:
\begin{equation}
	\begin{split}
		& \sum_{\alpha=0}^{\exPcOrder} \exPcCof{ \stochDisp }{\text{a}, \alpha }^v  \, \exPcMultiHermite{ \alpha}{ \setYoungRVs, \setMrmRVs, \setNoisRVs }  =  \sum_{\alpha=0}^{\exPcOrder} \exPcCof{ \stochDisp }{ \text{f},\alpha }^v  \, \exPcMultiHermite{ \alpha }{ \setYoungRVs, \setMrmRVs, \setNoisRVs } \\
		& + \kgain^v \bigg( \sum_{\alpha=0}^{\exPcOrder} \hat{\vecy}_{r,\alpha}^v  \, \exPcMultiHermite{\alpha}{\setYoungRVs, \setMrmRVs, \setNoisRVs} - \rhoCof^v \, \Hproj^v \, \sum_{\alpha=0}^{\exPcOrder} \exPcCof{ \stochDisp }{\text{f},\alpha }^v  \, \exPcMultiHermite{ \alpha }{ \setYoungRVs, \setMrmRVs, \setNoisRVs } - \sum_{\alpha=0}^{\exPcOrder} \exPcCof{\stochMrm}{\alpha}^v   \, \exPcMultiHermite{\alpha}{\setYoungRVs, \setMrmRVs, \setNoisRVs} - \sum_{\alpha=0}^{\exPcOrder} \exPcCof{\stochNoise}{\alpha}^v \, \exPcMultiHermite{\alpha}{\setYoungRVs, \setMrmRVs, \setNoisRVs} \bigg).
	\end{split}
	\label{eq:PC_representation_GMKF}
\end{equation}
Leveraging the linearity property, this expansion can be restructured so that the PC coefficients of the posterior displacement for $\nrep$ sets of observation data are directly obtained from
\begin{equation}
	\begin{split}
		\exPcCof{ \stochDisp }{\text{a}, \alpha }^v = \exPcCof{ \stochDisp }{ \text{f},\alpha }^v + \kgain^v \bigg(\frac{\sum_{\sclr = 1}^{\nrep}\hat{\vecy}_{r,\alpha}^v}{\nrep} - \rhoCof^v \, \Hproj^v \exPcCof{ \stochDisp }{ \text{f},\alpha }^v - \exPcCof{\stochMrm}{\alpha}^v - \exPcCof{\stochNoise}{\alpha}^v \bigg).
	\end{split}
	\label{eq:PC_coefficiants_GMKF}
\end{equation}
The PC-based GMKF's significant advantage is that the assimilated displacement's PC coefficients are obtained from purely algebraic means. \Rev{At the same time, we need to recall that the updated random variable only has the correct probability distribution in the linear Gaussian case and that nonlinear approximations of the conditional expectation are required in general, as outlined, for instance, in \cite{matthies2016parameter,hoang2023machine}. However, as our results will show, the linear filter can be fairly accurate, and since our focus is on the conceptual connection of the statFEM approach with Polynomial Chaos filtering, we omit discussing these more involved updating formulas.}

Finally, by back-substituting $\exPcCof{ \stochDisp }{\text{a}, \alpha }$ into \eqref{eq:extended_pc_expansions_assimilated_displacement} and sampling from the random variables  $\setYoungRVs$, $\setMrmRVs$, and $\setNoisRVs$, we can generate samples of the posterior displacement and compute empirical statistical moments, such as the mean value of the posterior displacement, covariance, skewness and kurtosis. The first two statistical moments - mean value and covariance of the posterior displacement - can also be determined using the orthogonality properties of Hermite polynomials as
\begin{equation}
	\begin{split}
		& \meanVec{\stochDisp_{\text{a}}^v}  = \exPcCof{ \stochDisp }{\text{a}, 0 }^v,\\
		& \covMat{\stochDisp_{\text{a}}^v} = \sum_{\alpha = 1}^{\exPcOrder} \exHermiteDobInner{\alpha}{\setYoungRVs}{\setMrmRVs}{\setNoisRVs} \, \exPcCof{ \stochDisp }{\text{a}, \alpha }^v \, (\exPcCof{ \stochDisp }{\text{a}, \alpha }^v)^\top,
	\end{split}
	\label{eq:Posterior_Mean_and_Cov_Disp_PC_Expansion}
\end{equation}
see \cite{ghanem2003stochastic}. With these results in hand, the mean of the true displacement $\meanVec{\trueDisp^v}$ and its covariance $\covMat{\trueDisp^v}$ at the sensor locations can be computed by
\begin{equation}
	\begin{split}
		& \meanVec{\trueDisp^v} = \Hproj^v \, \meanVec{\stochDisp_{\text{a}}^v},\\
		& \covMat{\trueDisp^v}  = \Hproj^v \, \covMat{\stochDisp_{\text{a}}^v} \, (\Hproj^v)^\top + \covMat{\stochMrm^v}.
	\end{split}
	\label{eq:Mean_and_Cov_True}
\end{equation}

\subsection{Identification of Hyperparameters}\label{sec:identify_hyperparameters}
The statistical model in \eqref{eq:statistical_generating_model3} relies on the dimension-specific unknown scale factor $\rhoCof^v$ and the unknown standard deviations of each mode $\mrmKlCof{m}$, which must be identified based on observational data prior to updating the PC coefficients of the displacement. Consequently, these parameters are treated as hyperparameters, defined as
\begin{equation}
	\hypVec^v = \big[\rhoCof^v, \,  \mrmKlCof{1}, \, \dots, \, \mrmKlCof{\nKlRvOfMrm} \big].
	\label{eq:hyperparameter_vector}
\end{equation}
In the remainder of this section, we will temporarily omit the dimension-specific notation $v$ to simplify the discussion. Note that the marginal likelihood defined in \eqref{eq:marginal_likelihood} is parameterized in the hyperparameters, i.e. $\PDF{\vecY}{\multipleObs} = \PDF{\vecY}{\multipleObs; \hypVec}$ as in \cite{girolami2021statistical,narouie2023inferring}. These hyperparameters are estimated by maximizing the marginal likelihood as
\begin{equation}
	\identifiedHypVec = \argmax_{\hypVec}{\Big( \PDF{\vecY}{\multipleObs; \hypVec} \Big)}.
	\label{eq:maximizing_marginl_likelihood}
\end{equation}
The identification process begins with a single dataset $\SingleObs$ and subsequently expands to multiple datasets $\multipleObs$, and the process is described in five stages as follows:\\

\noindent  \textbf{I. Statistical Model in Terms of $\setYoungRVs$:} We focus exclusively on the Gaussian component $\gaussianPartSingleObs$ of the statistical generating model outlined in \eqref{eq:statistical_generating_model2}, composed of the model-reality mismatch $\stochMrm$ and the stochastic noise $\stochNoise$, i.e.,
\begin{equation}
	\gaussianPartSingleObs = \underbrace{ \stochMrm(\setMrmRVs) }_{\text{Gaussian}} + \underbrace{ \stochNoise(\setNoisRVs) }_{\text{Gaussian}}.
	\label{eq:Gaussian_Part_statistical_generating_model2}
\end{equation}
By substituting it  into \eqref{eq:statistical_generating_model2}, we obtain
\begin{equation}
	\surrSingleObs = \vecg(\setYoungRVs) + \gaussianPartSingleObs
	\label{eq:hyp2_revised},
\end{equation}
where $\vecg(\setYoungRVs) = \rhoCof \, \Hproj \, \stochDisp_{h}^v(\setYoungRVs)$.\\

\noindent  \textbf{II. Joint PDF of Forecasted $\surrSingleObs$ and $\vecg$ :} The PDF of the forecasted observation $\surrSingleObs$ is then characterized by
\begin{equation}
	\PDF{\surrSingleObs, \vecg}{\surrSingleObs, \vecg} = \PDF{\surrSingleObs \, | \, \vecg}{\surrSingleObs \, | \, \vecg} \PDF{\vecg}{\vecg},
	\label{eq:joint_pdf}
\end{equation}
where the conditional PDF $\PDF{\surrSingleObs \, | \, \vecg}{\surrSingleObs \, | \, \vecg}$ is defined as
\begin{equation}
	\PDF{\surrSingleObs \, | \, \vecg}{\surrSingleObs \, | \, \vecg} = \underbrace{\frac{1}{\sqrt{(2\pi)^{\nsen} \det{\covMat{\stochMrm + \stochNoise}}}}}_{\beta(\setMrmKlCofs)} \exp{-\frac{1}{2} \bigg[\Big(\surrSingleObs - \vecg\Big)^\top \covMat{\stochMrm + \stochNoise}^{-1} \Big(\surrSingleObs - \vecg\Big) \bigg]}
	\label{eq:conditional_pdf_of_set_young_rvs},
\end{equation}
and $\covMat{\stochMrm + \stochNoise} =  \covMat{\stochMrm}(\setMrmKlCofs) + \covMat{\stochNoise}$ is the combined noise and model-reality mismatch covariance matrix.\\

\noindent  \textbf{III. Marginal PDF of Forecasted $\surrSingleObs$ :} The marginal likelihood of the forecasted observation $\surrSingleObs$ is then given as
\begin{equation}
	\PDF{\surrSingleObs}{\surrSingleObs} = \int \PDF{\surrSingleObs \, | \, \vecg}{\surrSingleObs \, | \, \vecg} \PDF{\vecg}{\vecg} \, d\vecg,
	\label{eq:marginal_likelihood_forcasted_observation}
\end{equation}
which can be expressed, because of $\PDF{\vecg}{\vecg} \, d\vecg = \PDF{\vecu_{\text{f}}}{\vecu} \, d\vecu$, as
\begin{equation}
	\begin{split}
		\PDF{\surrSingleObs}{\surrSingleObs} =  \int \PDF{\surrSingleObs \, | \, \vecg}{\surrSingleObs \, | \, \vecg(\vecu)} \PDF{\vecu_{\text{f}}}{\vecu} \, d\vecu,
	\end{split}
	\label{eq:marginal_likelihood_forcasted_observation_change_PDF}
\end{equation}
where $\vecg(\vecu) = \rhoCof \, \Hproj \, \vecu$ with a slight abuse of notation. \\

\noindent  \textbf{IV.  Transformation from $\vecu$ to $\setYoungRVs$:} Calculating the marginal likelihood in \eqref{eq:marginal_likelihood_forcasted_observation_change_PDF} requires the evaluation of the integral stated above. However, $\PDF{\vecu_{\text{f}}}{\stochDisp}$ is generally unknown, which makes the integration of \eqref{eq:marginal_likelihood_forcasted_observation_change_PDF} unfeasible. One possible solution is to assume $\PDF{\vecu_{\text{f}}}{\stochDisp}$ to be a multivariate Gaussian distribution with the mean and covariance from \eqref{eq:Mean_and_Cov_Disp_PC_Expansion}. In this case, the integration can be achieved using Monte Carlo sampling from the multivariate Gaussian distribution, which is parameterized in the unknown hyperparameters $\hypVec$. Herein,   making use of \eqref{eq:statistical_generating_model3}, we consider an alternative PC approximation where the surrogate observation data $\surrSingleObs$ is expressed as the sum of a polynomial function of $\setYoungRVs$ and the Gaussian term $\gaussianPartSingleObs$. For this purpose, we replace $\stochDisp$ with its PC expansion from \eqref{eq:Disp_PC_Expansion} with its germ distribution $\PDF{\setYoungRVs}{\setYoungRVs}  \,  \mathrm{d} \setYoungRVs$, with $\PDF{\setYoungRVs}{\setYoungRVs}$ representing the standard multivariate Gaussian distribution, see e.g. \cite{zhu2023stochastic}.  Substituting \eqref{eq:conditional_pdf_of_set_young_rvs} in to \eqref{eq:marginal_likelihood_forcasted_observation}, we obtain
\begin{equation}
	\PDF{\surrSingleObs}{\surrSingleObs} \approx \frac{\beta(\setMrmKlCofs)}{\sqrt{(2\pi)^{\pcOrderDisp} }} \int \exp{-\frac{1}{2} \bigg[\Big(\surrSingleObs - \vecg(\setYoungRVs)\Big)^\top \covMat{\stochMrm + \stochNoise}^{-1} \Big(\surrSingleObs - \vecg(\setYoungRVs)\Big) \bigg]} \exp{-\frac{1}{2} \setYoungRVs^\top \setYoungRVs} \, d\setYoungRVs
	\label{eq:marginal_likelihood_forcasted_observation_revised}
\end{equation}
which describes the forecast distribution of $\surrSingleObs$ and is a function of the statistical model uncertainty from $\vecu$, $\vecd$ and $\vece$. On the other hand, the experimental data $\SingleObs$ is the observed outcome from the physical system, which we are trying to validate against the forecasted $\surrSingleObs$. Therefore, the marginal PDF of a single $\SingleObs$ is given as
\begin{equation}
	\PDF{\surrSingleObs}{\SingleObs} \approx \int \PDF{\surrSingleObs \, | \, \vecg}{\SingleObs \, | \, \vecg(\setYoungRVs)} \PDF{\setYoungRVs}{\setYoungRVs} \, d\setYoungRVs,
	\label{eq:marginal_likelihood_for_observed_data}
\end{equation}
and for multiple sets of observation data $\multipleObs$, we have
\begin{equation}
	\begin{split}
		\PDF{\surrSingleObs}{\multipleObs; \hypVec } & = \PDF{\surrSingleObs}{\multipleObs} \approx \int \prod_{r=1}^{\nrep} \PDF{\surrSingleObs \, | \, \vecg}{\SingleObs \, | \, \vecg(\setYoungRVs)} \PDF{\setYoungRVs}{\setYoungRVs} \, d\setYoungRVs\\
		& = \frac{\beta^{\nrep}(\setMrmKlCofs)}{\sqrt{(2\pi)^{\pcOrderDisp} }} \int \exp{-\frac{1}{2}  \sum_{\sclr = 1}^{\nrep} \bigg[\Big(\SingleObs - \vecg(\setYoungRVs)\Big)^\top \covMat{\stochMrm + \stochNoise}^{-1} \Big(\SingleObs - \vecg(\setYoungRVs)\Big) \bigg]} \exp{-\frac{1}{2} \setYoungRVs^\top \setYoungRVs} \, d\setYoungRVs.
	\end{split}
	\label{eq:marginal_likelihood_for_multiple_observed_data}
\end{equation}\\

\noindent \textbf{V. Numerical Computation of $\PDF{\surrSingleObs}{\multipleObs}$:} Calculating the marginal likelihood \eqref{eq:marginal_likelihood_for_multiple_observed_data} requires the evaluation of the integral. This integration can be accomplished numerically with $\nGL$ number of Gauss-Legendre quadrature points similarly to \eqref{eq:PC_Coof_identify}, so that
\begin{equation}
	\PDF{\surrSingleObs}{\multipleObs; \hypVec}  \approx \frac{\beta^{\nrep}(\setMrmKlCofs)}{\sqrt{(2\pi)^{\nGL}}} \sum_{\scln = 1}^{\nGL} \exp{-\frac{1}{2} \sum_{\sclr = 1}^{\nrep} \bigg[ \Big(\SingleObs - \vecg(\setYoungRVs_n) \Big)^\top \covMat{\stochMrm + \stochNoise}^{-1} \Big(\SingleObs - \vecg(\setYoungRVs_n) \Big) \bigg]} \wGL.
	\label{eq:hyp7}
\end{equation}
Moreover, maximizing the natural logarithm of the marginal likelihood is preferable to improving numerical stability. We minimize the negative log marginal likelihood $\vartheta(\hypVec)$, i.e.,
\begin{equation}
	\identifiedHypVec = \argmax_{\hypVec}{\Big( \ln{\PDF{\surrSingleObs}{\multipleObs; \hypVec}}\Big)} = \argmin_{\hypVec} \, \vartheta(\hypVec)
	\label{eq:minMarginal}
\end{equation}
with the function $\vartheta(\hypVec)$ defined as follows:
\begin{equation}
	\begin{split}
		\vartheta(\hypVec) & = \frac{\nrep}{2}\ln{\det{\covMat{\stochMrm + \stochNoise}} } + \frac{\nsen \nrep}{2}\ln{2\pi}  -                                                                                                                                                                     \\
		& \ln{\sum_{\scln = 1}^{\nGL} \exp{\underbrace{-\frac{1}{2} \sum_{\sclr = 1}^{\nrep} \bigg[ \Big(\SingleObs - \vecg(\setYoungRVs_n) \Big)^\top \covMat{\stochMrm + \stochNoise}^{-1} \Big(\SingleObs - \vecg(\setYoungRVs_n) \Big) \bigg]}_{\scly^{\star}_\scln}} \wGL }.
	\end{split}
	\label{eq:log_marginal_likelihood_theta_func}
\end{equation}
To prevent overflows and maintain accuracy when dealing with large exponentials, we slightly modified the shifting technique proposed in \cite{blanchard2021accurately}. The weighted log-sum-exp $\ln{\sum_{\scln = 1}^{\nGL} \wGL \exp{\bullet}}$, the log-sum-exp of a vector $\{\scly^{\star}_\scln\}_{n=1}^{\nGL}$  with weights $\wGL$, ensures numerical stability by incorporating a shifting technique based on the maximum value of $\{\scly^{\star}_\scln\}_{n=1}^{\nGL}$. Therefore, \eqref{eq:log_marginal_likelihood_theta_func} can be reformulated as
\begin{equation}
	\vartheta(\hypVec) = \frac{\nrep}{2}\ln{\det{\covMat{\stochMrm + \stochNoise}}} + \frac{\nsen \nrep}{2}\ln{2\pi}  - \scly^{\star}_{\text{max}} - \ln{ \sum_{\substack{\scln = 1 \\ \scln \neq \sclk} }^{\nGL} \wGL \exp{\scly_\scln^{\star}- \scly^{\star}_{\text{max}}}},
	\label{eq:shifting_algorithm}
\end{equation}
where $\scly^{\star}_{\text{max}} = \max{\{\scly^{\star}_\scln\}_{n=1}^{\nGL}}$. To enhance numerical stability further, the hyperparameters $\mrmKlCof{\sclm}$ are transformed into natural logarithmic space $\ln{\mrmKlCof{\sclm}}$ during the minimization process. Additionally, the determinant and inversion of $\covMat{\stochMrm} + \covMat{\stochNoise}$ are computed using a Cholesky decomposition.

\section{Numerical Examples}\label{sec:Numerical_Examples}
This section presents a series of numerical examples to illustrate the application and effectiveness of the proposed methods. We consider three distinct test cases.

\begin{enumerate}
	\renewcommand{\labelenumi}{\Roman{enumi}.}
	\item The initial example \autoref{subsec:One-dimensional_Tension_Bar} involves a $1$D tension bar characterized by a linear elastic material model and uncertain Young's modulus. Here, we aim to re-identify predefined hyperparameters with several numbers of hyperparameters, specifically within the range $\nKlRvOfMrm\in\left\{2,3,4,9\right\}$, and report the accuracy of the re-identification.

	\item The second example \autoref{subsec:one_dimensional_tension_bar_with_inhomogeneous_young_modulus} concerns the same $1$D tension bar but introduces synthetic observational data resulting from a heterogeneous Young's modulus. This scenario is designed to test the ability of the proposed model-reality mismatch term to capture a model error that arises from a typical engineering assumption.

	\item The final example \autoref{subsec:two_dimensional_infinite_plate_with_a_hole_and_heterogeneous_young_s_modulus} extends the method to a $2$D scenario involving a plate with a hole. This test aims to demonstrate the capability of model-reality mismatch to handle discrepancies in Young's modulus between simulation and reality, particularly in areas surrounding the hole, which may arise due to the drilling or stamping of the hole in the plate.
\end{enumerate}

\subsection{One-Dimensional Tension Bar with Homogeneous Young's Modulus}\label{subsec:One-dimensional_Tension_Bar}

In this example, we examine a $1$D tension bar subject to a point load at the tip.\par

\begin{figure}[!htb]
	\centering
	\includegraphics{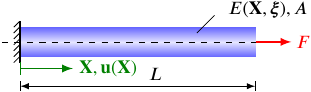}
	\caption{One-Dimensional tension bar.}
	\label{fig:one-dimensional_Tension_Bar}
\end{figure}

\noindent\textbf{Prior model:} The prior displacement is assumed to stem from a boundary value problem with an uncertain Young's modulus, which is modeled as a weakly stationary random field. The geometry and the loading on the bar under tension are illustrated in \autoref{fig:one-dimensional_Tension_Bar}, and the stochastic boundary value problem is described as
\begin{equation}
	\begin{split}
		-\frac{\partial}{\partial \spatialPoints} & \bigg[E(\spatialPoints, \setYoungRVs) \, A \, \frac{\partial \stochDisp(\spatialPoints, \setYoungRVs)}{\partial \spatialPoints }\bigg]   = \vec0, \quad \spatialPoints \in [0,L]                            \\
		\stochDisp(\spatialPoints, \setYoungRVs)          & \bigg|_{\spatialPoints=0} = \vec0, \,\,\,\,\,\,\,\, \frac{\partial \stochDisp(\spatialPoints, \setYoungRVs)}{\partial \spatialPoints}\bigg|_{\spatialPoints=L} = \frac{F}{A \, E(\spatialPoints, \setYoungRVs)} \bigg|_{\spatialPoints=L},
	\end{split}
	\label{eq:1D_bar_PDE}
\end{equation}
where $\sclL$ is the length of the bar, $E(\spatialPoints, \setYoungRVs)$ the uncertain Young's modulus, $A$ the cross-sectional area, $\stochDisp(\spatialPoints, \setYoungRVs)$ the stochastic displacement, $\sclF$ the concentrated load at the tip of the bar. Moreover, the Dirichlet and Neumann boundary conditions are also given in \eqref{eq:1D_bar_PDE}.

\begin{figure}[!ht]
    \vspace{-0.5cm}
	\centering
	\subfloat[]{
		\centering
		\includegraphics{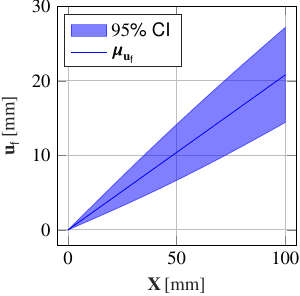}
		\label{subfig:one-dimensional_tension_bar_mean_u_ci_u}}
	\hspace*{1cm}
	\subfloat[]{
		\centering
		\includegraphics{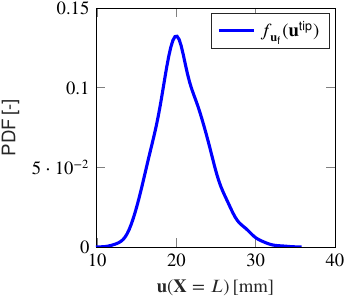}
		\label{subfig:KDE_u_at_tip}}
	\caption{\textbf{Prior Displacement of $1$D Tension Bar}: (a) Mean of prior stochastic displacement $\meanVec{\stochDisp_{\text{f}}}$ (blue line) and its $95\%$ credible interval (CI) (shaded area); (b) PDF of stochastic displacement at tip of the bar $\PDF{\vecu_{\text{f}}}{\stochDisp^{\text{tip}}}$.}
	\label{fig:1D_bar_prior}
\end{figure}

The uncertainty in Young's modulus is considered to follow a log-normally distributed weakly stationary random field with a mean $\meanFunc[E]{\spatialPoints} = \meanOf{E} = 200 \,\si{\GPa}$ and a standard deviation $\stdFunc[E]{\spatialPoints} = \stdOf{E} = 20 \,\si{\GPa}$. It is important to note that terms such as stationary, weakly stationary, and non-stationary are used in describing the probabilistic characteristics of a random field as described in \ref{subsec:Definition_Of_Random_Variables_And_Random_Fields}. On the other hand, terms like homogeneous and heterogeneous are used in the context of solid mechanics. A homogeneous material has uniform properties throughout the domain, while a heterogeneous material shows variability. Using \eqref{eq:mean_and_std_of_kappa_based_on_E}, the parameters $\meanOf{\kappa} = 5.2787$ and $\stdOf{\kappa} = 0.1980$ of the weakly stationary Gaussian random field $\gaussRFs$ are obtained. The Gaussian random field $\gaussRFs$ is endowed with the squared exponential kernel with a correlation length $\oneDcorlength{\kappa} = L/10$. Therefore, the number of KL terms, derived from \eqref{eq:explained_variance},  is determined to be $\nKlRvOfYoung = 10$. The bar's length is $\sclL = 100 \,\si{\mm}$, the cross-sectional area is $A = 20 \, \si{\mm^2}$ and the force at the tip of the bar is $\sclF=800 \, \si{kN}$. Considering equation \eqref{eq:POrder} with a univariate polynomial order $\sclp = 2$, the number of PC coefficients is calculated to be $\pcOrderDisp = 66$.  The PC coefficients of the stochastic displacement $\stochDisp_{h} (\setYoungRVs) \in \bbR^{100 \times 1}$ are then identified using \eqref{eq:PC_Coof_identify}. In \autoref{subfig:one-dimensional_tension_bar_mean_u_ci_u}, we show the prior mean $\meanVec{\stochDisp_{\text{f}}}$ of the stochastic displacement and its $95\%$ credible interval (CI) and the PDF of the stochastic displacement at the tip of the bar $\PDF{\vecu_{\text{f}}}{\stochDisp^{\text{tip}}}$ is depicted in \autoref{subfig:KDE_u_at_tip}. \\

\noindent\textbf{Data generation:} To generate synthetic observation data $\SingleObs$, we utilize the analytical solution for the displacement, which is represented by the formula
\begin{equation}
	\vecu_{\text{ho}}(\spatialPoints) = \frac{\sclF}{A  \meanOf{E}}\spatialPoints.
	\label{eq:disp_hom}
\end{equation}
To get a cleaner observational dataset without the nodal discretization error inherent in FEM analysis, we decided to use the analytical solution to generate the observational data, and the prior was calculated using the FEM. To this end, we choose $\nsen = 9$ and also introduce observational noise with a standard deviation defined for each sensor as $\noisPcCof{1}, \dots, \noisPcCof{k}, \dots, \noisPcCof{\nsen} = 0.1$, where $v = 1$.

Additionally, to account for the model-reality mismatch, we incorporate predefined hyperparameters. These predefined hyperparameters are used to generate different sets of synthetic observation data and serve as benchmarks to evaluate the accuracy of the identification process described in \autoref{sec:identify_hyperparameters}. We adjust the correlation length of the model-reality mismatch $\oneDcorlength{\stochMrm}$ to effectively manipulate the number of hyperparameters with the help of \eqref{eq:explained_variance}. We choose $\oneDcorlength{\stochMrm}= \sclL /10 = 10 \,\si{\mm}$, therefore, the number of KL terms is $\nKlRvOfMrm = 9$. \\

\noindent\textbf{Hyperparameter identification:} The results of the identified hyperparameters for this case are illustrated in \autoref{tab:identified_hyperparameters_Md_9} and for other cases can be found in \autoref{Appendix:Identified_Hyperparameters_for_Example_1}. The table presents the results from identifying the hyperparameters $\identifiedHypVec$ with different numbers of sensor readings $\nrep$.

\begin{table}[H]
	\centering
	\begin{tabular}{cccccc}
		\toprule
		                     & Predefined & $\nrep = 1$      & $\nrep = 10$     & $\nrep = 100$    & $ \nrep = 1000$  \\\midrule
		$ \rhoCof $          & $ 1.5 $    & $ 1.537015 $     & $ 1.700901 $     & $ 1.646780 $     & $ 1.651428 $     \\
		$ \mrmKlCof{1} $     & $ 3.0 $    & $ 5.446338 $     & $ 3.688666 $     & $ 3.400279 $     & $ 3.073303 $     \\
		$ \mrmKlCof{2} $     & $ 3.0 $    & $ 3.189917 $     & $ 3.215770 $     & $ 3.019729 $     & $ 2.976911 $     \\
		$ \mrmKlCof{3} $     & $ 2.5 $    & $ 1.055175 $     & $ 2.532338 $     & $ 2.507561 $     & $ 2.517128 $     \\
		$ \mrmKlCof{4} $     & $ 2.3 $    & $ 0.849787 $     & $ 2.099813 $     & $ 2.345987 $     & $ 2.332457 $     \\
		$ \mrmKlCof{5} $     & $ 2.2 $    & $ 0.792093 $     & $ 1.801989 $     & $ 2.194500 $     & $ 2.184333 $     \\
		$ \mrmKlCof{6} $     & $ 0.7 $    & $ 0.478803 $     & $ 0.396637 $     & $ 0.618122 $     & $ 0.709982 $     \\
		$ \mrmKlCof{7} $     & $ 0.4 $    & $ 0.004204 $     & $ 0.327544 $     & $ 0.360578 $     & $ 0.416197 $     \\
		$ \mrmKlCof{8} $     & $ 0.3 $    & $ 0.001100 $     & $ 0.291332 $     & $ 0.315157 $     & $ 0.316627 $     \\
		$ \mrmKlCof{9} $     & $ 0.2 $    & $ 0.000482 $     & $ 0.252314 $     & $ 0.204664 $     & $ 0.213573 $     \\
		\bottomrule
		$ \Rev{err_{1000}} $ &            & $ \Rev{2.2302} $ & $ \Rev{2.4469} $ & $ \Rev{0.6957} $ & $ \Rev{0.0623} $ \\
		\bottomrule
	\end{tabular}
	\caption{Identified hyperparameters with $ \isotropicTwoDcorlength{ \stochMrm } = 10 \,\si{\mm}$ \label{tab:identified_hyperparameters_Md_9}}
\end{table}

\begin{figure}[!ht]
    \vspace{-0.5cm}
	\centering
	\subfloat[]{
		\centering
		\includegraphics{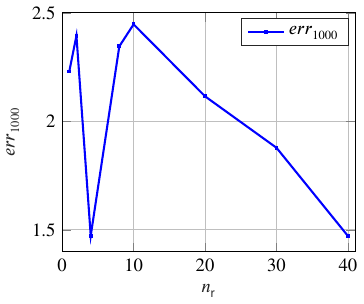}
		\label{subfig:err_nr_1_8}}
	\hspace*{1cm}
	\subfloat[]{
		\centering
		\includegraphics{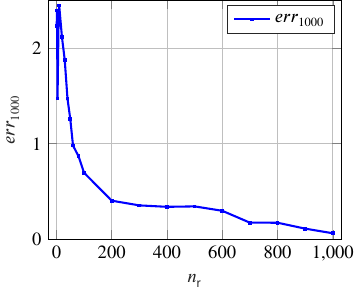}
		\label{subfig:err_nr}}
	\caption{\Rev{Absolute error \eqref{eq:err1000} in the range (a) $\nrep = [1, 40]$ and (b) $\nrep = [1, 1000]$.}}
	\label{fig:err_nr_convergence}
\end{figure}

As the number of repetitions $\nrep$ increases from $1$ to $1000$, the identified hyperparameters $\identifiedHypVec$ converge closer to the predefined values. More data, i.e., higher $\nrep$, enable a more accurate identification process. Specifically based on \autoref{tab:identified_hyperparameters_Md_9}, for the hyperparameters like $\mrmKlCof{1}$, $\mrmKlCof{2}$, and $\mrmKlCof{3}$, the identified values with $\nrep = 100$ and $\nrep = 1000$ are notably closer to the predefined values compared to $\nrep = 1$. Lower hyperparameters such as $\mrmKlCof{8}$, $\mrmKlCof{9}$ initially show very low values with $\nrep = 1$ but approach their predefined values as $\nrep$ increases. This indicates that these hyperparameters are only properly captured with a sufficiently high $\nrep$. \Rev{Note that the identified coefficients $\mrmKlCof{i}$ may not be uniquely (re)-identifiable, and thus errors in their (re)-identification provide limited insight into the accuracy of the considered statFEM approach. Therefore, we define an absolute error as}
\begin{equation}\label{eq:err1000}
	\Rev{err_{1000} = \norm{\meanVec{\trueDisp} - \meanVec{\multipleObs_{1000}}},}
\end{equation}
\Rev{which represents the norm of the error between the inferred true displacement from statFEM, as given in \eqref{eq:Mean_and_Cov_True}, and the mean of $1000$ realizations of the data, $\meanVec{\multipleObs_{1000}}$. The convergence study is shown in \autoref{fig:err_nr_convergence}. Although fluctuations are observed for $\nrep=1$ to $\nrep=10$ (\autoref{subfig:err_nr_1_8}), the error norm decreases as the number of sensor readings increases, as shown in \autoref{subfig:err_nr}.
}

\begin{figure}[!ht]
	\centering
	\hspace*{-0.9cm}
	\subfloat[$\nrep = 1$]{
		\centering
		\includegraphics{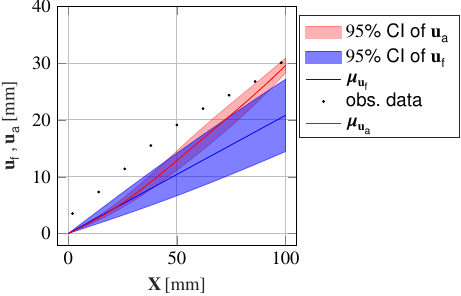}
		\label{subfig:tensionBar_u_y_Md6_nSen9_nRed1}}
	\subfloat[$\nrep = 10$]{
		\centering
		\includegraphics{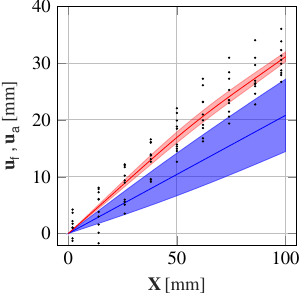}
		\label{subfig:tensionBar_u_y_Md6_nSen9_nRed10}}
	\\
	\hspace*{-0.9cm}
	\subfloat[$\nrep = 100$]{
		\centering
		\includegraphics{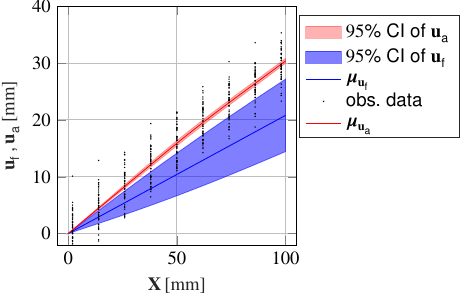}
		\label{subfig:tensionBar_u_y_Md6_nSen9_nRed50}}
	\subfloat[$\nrep = 1000$]{
		\centering
		\includegraphics{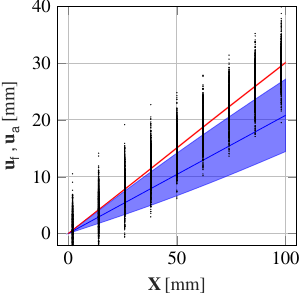}
		\label{subfig:tensionBar_u_y_Md6_nSen9_nRed1000}}
	\caption{\textbf{Posterior Displacement of $1$D Tension Bar}: The blue line is the mean of prior stochastic displacement $\meanVec{\stochDisp_{\text{f}}}$ and its $95\%$ credible interval (CI). The black dots are the observation data. The red line represents the mean of the prior stochastic displacement $\meanVec{\stochDisp_{\text{a}}}$ and the shaded red area denotes its $95\%$ CI.}
	\label{fig:tensionBar_u_y_Md6_nSen9}
\end{figure}

Once the hyperparameters are identified, the covariance matrix of the model-reality mismatch $\covMat{\stochMrm}$ can be determined using \eqref{eq:Cov_Mrm_KL_Expansion}, and the K{\'a}lm{\'a}n gain $\kgain$ can be derived from \eqref{eq:kalman_gain}. By substituting the K{\'a}lm{\'a}n gain into PC-Based GMKF \eqref{eq:PC_coefficiants_GMKF}, we update the PC coefficients of the prior displacement $\exPcCof{ \stochDisp }{ f,\alpha }$ to compute the posterior PC coefficients $\exPcCof{ \stochDisp }{ a,\alpha }$ of the stochastic displacement. This methodology allows for straightforward computation of the posterior displacement's mean value $\meanVec{\stochDisp_{\text{a}}}$ and covariance $\covMat{\stochDisp_{\text{a}}}$ using \eqref{eq:Posterior_Mean_and_Cov_Disp_PC_Expansion}. The posterior results for $\nsen =9$ are depicted in \autoref{fig:tensionBar_u_y_Md6_nSen9}. As can be seen, the credible interval of the stochastic displacement $\meanVec{\stochDisp_{\text{a}}}$ is narrower than the prior interval $\meanVec{\stochDisp_{\text{f}}}$ and the more readings of observation data are used for inference, the narrower the credible interval becomes.

\begin{figure}[!htb]
	\centering
	\includegraphics{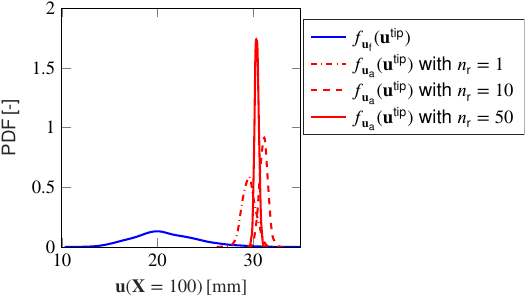}
	\caption{\textbf{PDF of stochastic displacement at tip of the bar}: The blue line is the PDF of prior stochastic displacement $\PDF{\vecu_{\text{f}}}{\stochDisp^{\text{tip}}}$, the red lines are the PDF of posterior stochastic displacement at tip of the bar $\PDF{\vecu_{\text{a}}}{\stochDisp^{\text{tip}}}$ for $\nrep = [1, \, 10, \, 50]$. For $\nrep=50$, the credible interval narrows significantly.}
	\label{fig:PDF_u_tip_Md4_nSen33_nRed50}
\end{figure}

By reinserting the updated PC coefficients $\exPcCof{ \stochDisp }{a, \alpha }$ into equation \eqref{eq:extended_pc_expansions_assimilated_displacement}, and then sampling from the sets of random variables $\setYoungRVs$, $\setMrmRVs$, and $\setNoisRVs$, we can generate a range of samples representing the posterior displacement. To evaluate the distribution of these displacement samples at the tip, we apply \textit{Kernel Density Estimation} (KDE) \cite{jones1993simple}, which is a non-parametric method to estimate the PDF of the posterior displacement at the tip of the bar, i.e., $\PDF{\vecu_{\text{a}}}{\stochDisp^{\text{tip}}}$. The estimated PDFs are derived for different sets of sensor readings, specifically for $\nrep =1$, $\nrep = 10$, and $\nrep = 50$ and shown in \autoref{fig:PDF_u_tip_Md4_nSen33_nRed50}.

\subsection{One-Dimensional Tension Bar with Inhomogeneous Young's Modulus}
\label{subsec:one_dimensional_tension_bar_with_inhomogeneous_young_modulus}
This section aims to examine if the model-reality mismatch, viewed as a stochastic process, can account for discrepancies in displacement caused by an inhomogeneous Young's modulus.\\

\noindent \textbf{Prior model:} The prior stochastic displacements are simulated with FEM and PCE based on the assumption of a homogeneous Young's modulus. To accomplish this, Young's modulus is sampled from a log-normally distributed random field, as described in \autoref{subsec:One-dimensional_Tension_Bar}.\\

\begin{figure}[!htb]
    \vspace{-0.5cm}
	\centering
	\hspace*{-0.5cm}
	\subfloat[Inhomogeneous Young's modulus]{
		\centering
		\includegraphics{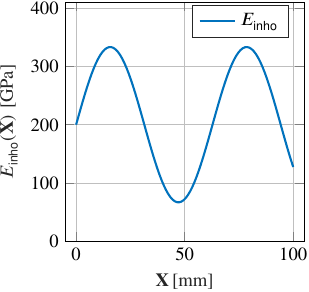}
		\label{subfig:E_inho_X}}
	\hspace*{1.5cm}
	\centering
	\subfloat[Displacement based on Inhomogeneous Young's modulus]{
		\includegraphics{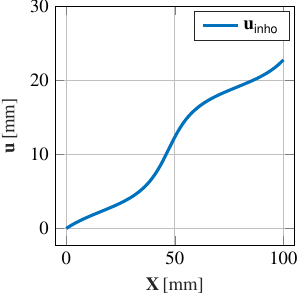}
		\label{subfig:U_inho}}
	\caption{Inhomogeneous Young's modulus and the corresponding displacement.}
	\label{fig:Inhomogeneous Young's Modulus and the Corresponding Displacement}
\end{figure}

\noindent\textbf{Data generation:} The experimental data are generated using an inhomogeneous Young's modulus, which is modeled using the sine function, shown in \autoref{subfig:E_inho_X}and described by
\begin{equation}\label{eq:local_young_modulus_function}
	E(X) = \meanOf{E} \Big(\frac{3\,\sin{\frac{X}{10}}}{2}  + 1 \Big).
\end{equation}
The displacement response, resulting from this inhomogeneous Young's modulus $\vecu_{\text{inh}}$, is obtained using FEM and illustrated along the bar's axis in \autoref{subfig:U_inho}.
Synthetic experimental data are generated by simulating displacements due to an inhomogeneous Young's modulus, with a predefined $\rhoCof = 1.3$. Gaussian noise, characterized by a standard deviation of $\sigma_{\stochNoise,\sclk} = 0.5$, is applied to each sensor, indexed by $\sclk = 1, \dots, \nsen$. Observation data are generated for two different sensor configurations, specifically with $\nsen = 10$ and $\nsen = 20$. \\

\begin{figure}[!htb]
	\vspace*{-0.5cm}
	\centering
	\hspace*{-0.9cm}
	\subfloat[$\nsen =20$, $\nrep = 1$]{
		\centering
		\includegraphics{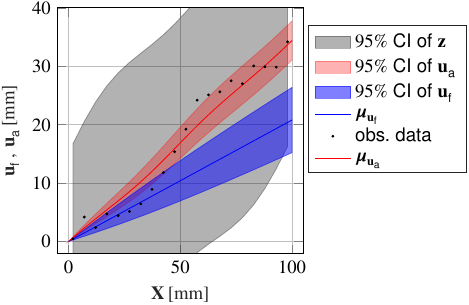}
		\label{subfig:inho_tensionBar_u_y_Md10_nSen20_nRed1}}
	\subfloat[$\nsen =20$, $\nrep = 10$]{
		\centering
		\includegraphics{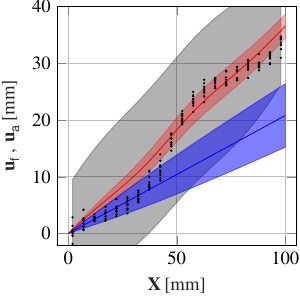}
		\label{subfig:inho_tensionBar_u_y_Md10_nSen20_nRed10}}
	\\
	\hspace*{-0.9cm}
	\subfloat[$\nsen =20$, $\nrep = 100$]{
		\centering
		\includegraphics{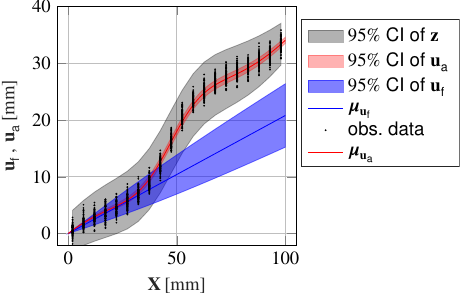}
		\label{subfig:inho_tensionBar_u_y_Md10_nSen20_nRed100}}
	\subfloat[$\nsen =20$, $\nrep = 1000$]{
		\centering
		\includegraphics{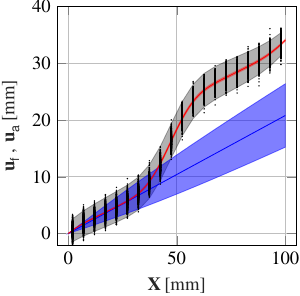}
		\label{subfig:inho_tensionBar_u_y_Md10_nSen20_nRed1000}}
	\caption{\textbf{Posterior Displacement of $1$D Tension Bar from Inhomogeneous Young's Modulus with $\nsen = 20$:} The blue line represents $\meanVec{\stochDisp_{\text{f}}}$, and the blue shaded area is the $95\percent$ of CI. The black dots are the generated observation data. The red lines represented $\meanVec{\stochDisp_{\text{a}}}$, and the shaded red area denotes its $95\percent$ CI. The black shaded area is the $95\percent$ CI of true displacement $\trueDisp$.}
	\label{fig:inho_tensionBar_u_y_Md10_nSen20}
\end{figure}

\noindent\textbf{Hyperparameter identification:} The next step involves identifying hyperparameters based on \autoref{sec:identify_hyperparameters}, and updating prior PC coefficients of stochastic displacement $\exPcCof{ \stochDisp }{ f,\alpha }$ according to  \eqref{eq:Posterior_Mean_and_Cov_Disp_PC_Expansion}. The results of these updates are illustrated in \autoref{fig:inho_tensionBar_u_y_Md9_nSen10} and \autoref{fig:inho_tensionBar_u_y_Md10_nSen20} for $\nrep = 10$ and $\nrep = 20$, respectively. Upon analyzing the figures, the most significant observations can be expressed as follows:
\begin{figure}[!htb]
	\vspace*{-0.5cm}
	\centering
	\hspace*{-0.9cm}
	\subfloat[$\nsen =10$, $\nrep = 1$]{
		\centering
		\includegraphics{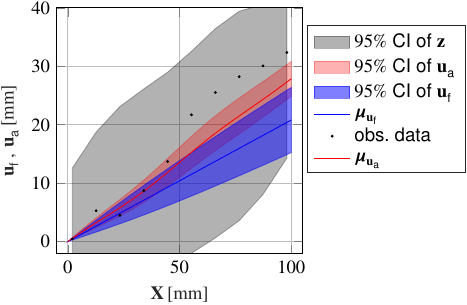}
		\label{subfig:inho_tensionBar_u_y_Md9_nSen10_nRed1}}
	\subfloat[$\nsen =10$, $\nrep = 10$]{
		\centering
		\includegraphics{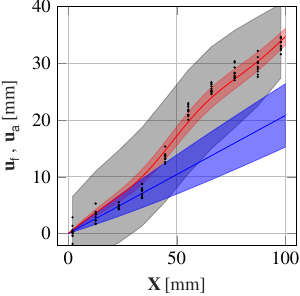}
		\label{subfig:inho_tensionBar_u_y_Md9_nSen10_nRed10}}
	\\
	\hspace*{-0.9cm}
	\subfloat[$\nsen =10$, $\nrep = 100$]{
		\centering
		\includegraphics{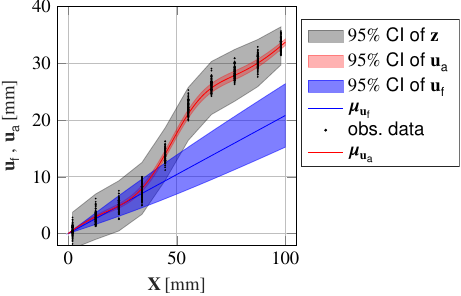}
		\label{subfig:inho_tensionBar_u_y_Md9_nSen10_nRed100}}
	\subfloat[$\nsen =10$, $\nrep = 1000$]{
		\centering
		\includegraphics{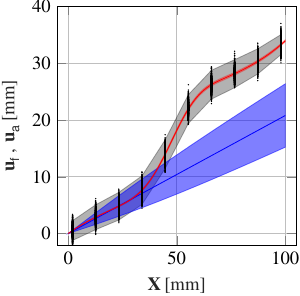}
		\label{subfig:inho_tensionBar_u_y_Md9_nSen10_nRed1000}}
	\caption{\textbf{Posterior displacement of $1$D tension bar from inhomogenous Young's modulus with $\nsen = 10$:} The blue line represents $\meanVec{\stochDisp_{\text{f}}}$, and the blue shaded area is the $95\percent$ of CI. The black dots are the generated observation data. The red lines represented $\meanVec{\stochDisp_{\text{a}}}$, and the shaded red area denotes its $95\%$ CI. The black shaded area is the $95\%$ CI of true displacement $\trueDisp$.}
	\label{fig:inho_tensionBar_u_y_Md9_nSen10}
\end{figure}

\begin{figure}[!htb]
	\center
	\includegraphics{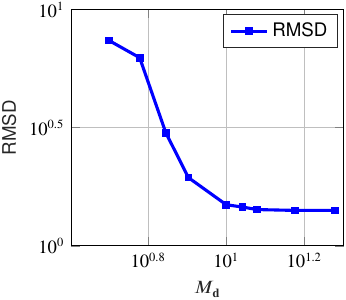}
	\caption{RMSD with $\nrep = 100$.}
	\label{fig:RMSD}
\end{figure}

\begin{enumerate}
	\renewcommand{\labelenumi}{\Roman{enumi}.}
	\item The results presented in \autoref{fig:inho_tensionBar_u_y_Md9_nSen10} were obtained using $10$ sensors, which uniformly distributed along the direction of the bar and given $\scll_{\vecd} = 25 \, \si{\mm}$. As shown in \autoref{subfig:inho_tensionBar_u_y_Md9_nSen10_nRed1}, \autoref{subfig:inho_tensionBar_u_y_Md9_nSen10_nRed10}, and \autoref{subfig:inho_tensionBar_u_y_Md9_nSen10_nRed100}, it is evident that as the number of sensor readings $\nrep$ increases, the mean value of the posterior displacement $\meanVec{\stochDisp_{\text{a}}}$ approaches the mean value of the observational data, and it's $95\percent$ credible interval narrows. The credible interval with $\nrep = 1000$ is extremely narrow and may not be distinguishable from the mean in \autoref{subfig:inho_tensionBar_u_y_Md9_nSen10_nRed1000}. The black shaded area represents the $95\percent$ credible interval of the true response $\trueDisp$.
	\item The results shown in \autoref{fig:inho_tensionBar_u_y_Md10_nSen20} were obtained using $\nsen = 20$  and $\scll_{\vecd} = 25 \, \si{\mm}$. As observed in \autoref{subfig:inho_tensionBar_u_y_Md10_nSen20_nRed1}, \autoref{subfig:inho_tensionBar_u_y_Md10_nSen20_nRed10}, \autoref{subfig:inho_tensionBar_u_y_Md10_nSen20_nRed100} and \autoref{subfig:inho_tensionBar_u_y_Md10_nSen20_nRed1000}, the increase in sensor readings leads to a posterior displacement aligning more closely with the mean value of the observational data. A comparison between \autoref{subfig:inho_tensionBar_u_y_Md9_nSen10_nRed1000}, which is derived from $\nsen = 10$ sensors, and \autoref{subfig:inho_tensionBar_u_y_Md10_nSen20_nRed1000}, obtained with $\nsen = 20$ sensors, reveals a smoother posterior displacement along the bar. It is evident that choosing more sensors enables the collection of more experimental data, thereby enhancing the quality of the inference.
\end{enumerate}
To assess the impact of the number of terms used in the expansion of model-reality mismatch, we conduct a convergence study focusing on the number of $\nKlRvOfMrm$ terms and the \textit{Root Mean Square Deviation} (RMSD), which is defined as

\begin{equation}
	\text{RMSD} = \sqrt{\dfrac{1}{\nMC} \sum_{i=1}^{\nMC}\Big(\Hproj \, \stochDisp_{ a } ( \setYoungRVs_i, \, \setMrmRVs_i, \, \setNoisRVs_i ) - \meanVec{\multipleObs} \Big)^2 }.
	\label{eq:RMSD}
\end{equation}
Here, $\setYoungRVs_i$ denotes the sampled stochastic degrees of freedom used in the forward problem, $\setMrmRVs_i$ represents the variables associated with the model-reality mismatch, and $\setNoisRVs_i$ corresponds to noise. The number of samples is represented by $\nMC$, set at $\nMC = 1000$. The Root Mean Square Deviation (RMSD) measures the discrepancy between the mean of the experimental data $\meanVec{\multipleObs}$ and the stochastic posterior displacement, where the transformation matrix $\Hproj$ projects posterior displacements $ \stochDisp_{ a }$ from node to sensor coordinates.  The convergence analysis shown in \autoref{fig:RMSD} indicates that the RMSD decreases as $\nKlRvOfMrm$ increases. It also reveals that the RMSD stabilizes when $\nKlRvOfMrm > 10$.

\begin{figure}[!htb]
	\vspace*{-0.5cm}
	\centering
	\hspace*{-0.9cm}
	\subfloat[$\nKlRvOfMrm = 5$]{
		\centering
		\includegraphics{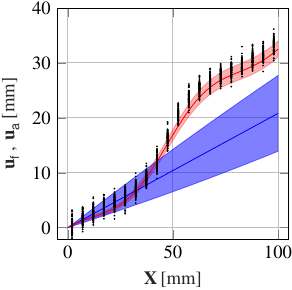}
		\label{subfig:con_inho_tensionBar_u_y_Md5_nSen20_nRed100}}
	\subfloat[$\nKlRvOfMrm = 10$]{
		\centering
		\includegraphics{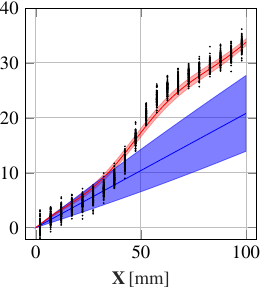}
		\label{subfig:con_inho_tensionBar_u_y_Md10_nSen20_nRed100}}
	\subfloat[$\nKlRvOfMrm = 15$]{
		\centering
		\includegraphics{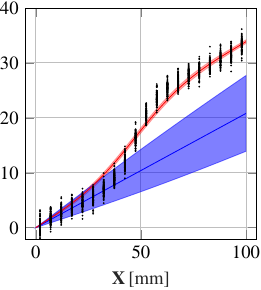}
		\label{subfig:con_inho_tensionBar_u_y_Md15_nSen20_nRed100}}
	\caption{\textbf{Posterior Displacement of $1$D Tension Bar for different $\nKlRvOfMrm$}: The blue line represents $\meanVec{\stochDisp_{\text{f}}}$, and the blue shaded area is the $95\percent$ of CI. The black dots are the generated observation data. The red lines represented $\meanVec{\stochDisp_{\text{a}}}$, and the shaded red area denotes its $95\percent$ CI. The legend of \autoref{fig:tensionBar_u_y_Md6_nSen9} is also applied here.}
	\label{fig:Conv_PF_tensionBar_u_y_ld50_Md24_ld10_Md42_ld1_Md50_nSen50_nRed100}
\end{figure}

\begin{figure}[!htb]
	\centering
	\includegraphics{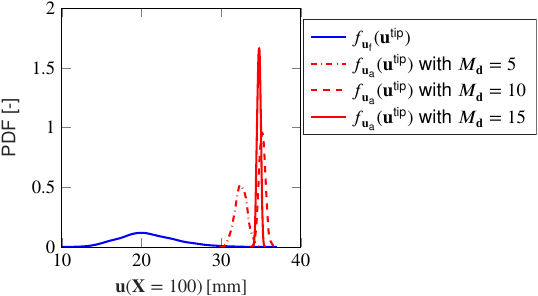}
	\caption{The convergence of PDF of displacement at the bar's tip for different $\nKlRvOfMrm$. A more detailed model-reality mismatch (with more KL parameters) can more accurately represent the experimental data.}
	\label{fig:PDF_Con_with_Md}
\end{figure}

The posterior displacement results for $\nKlRvOfMrm = [5, \, 10, \, 15]$ are displayed in \autoref{subfig:con_inho_tensionBar_u_y_Md5_nSen20_nRed100}, \autoref{subfig:con_inho_tensionBar_u_y_Md10_nSen20_nRed100} and \autoref{subfig:con_inho_tensionBar_u_y_Md15_nSen20_nRed100} respectively. It is observable that while the increase in $\nKlRvOfMrm$ does not significantly affect the mean value, the $95\percent$ confidence interval becomes narrower with higher values of $\nKlRvOfMrm$.
The PDF of the prior displacement $\PDF{\vecu_{\text{f}}}{\stochDisp^{\text{tip}}}$ and posterior $\PDF{\vecu_{\text{a}}}{\stochDisp^{\text{tip}}}$ at tip of the bar is shown in \autoref{fig:PDF_Con_with_Md}.

The analysis indicates that a more detailed model-reality mismatch can more accurately represent the experimental data, but only up to a specific limit, see \autoref{fig:RMSD}. More accurate representations can also be achieved when approximating the model-reality mismatch with other kernels, e.g., exponential kernel $ \corFunc[\kappa, \scls = \frac{1}{2}]{\spatialPoints}$, where analytical solutions of KL eigenpairs $\eigenPairs{i}$ are available. However, we have experienced a very slow convergence rate regarding the identification of hyperparameters even when $\nrep = 10$. Another challenging problem during optimization was the poor scaling of the starting hyperparameter.  When the starting values of the optimizer were very different from the solution, it caused slow convergence or even stopped the optimization. To address this, we tried many starting points for the optimizer to ensure convergence and subsequently used the optimized hyperparameters from one iteration as the starting point for the next. The numerical overflow from exponential terms in the objective function was reduced using the log-sum-exp method.

\subsection{Two-dimensional Plate with a Hole and Heterogeneous Young's Modulus} 
\label{subsec:two_dimensional_infinite_plate_with_a_hole_and_heterogeneous_young_s_modulus}
In this section, we examine a $2$D plate with a hole. \\

\begin{figure}[!htb]
	\centering
	\includegraphics{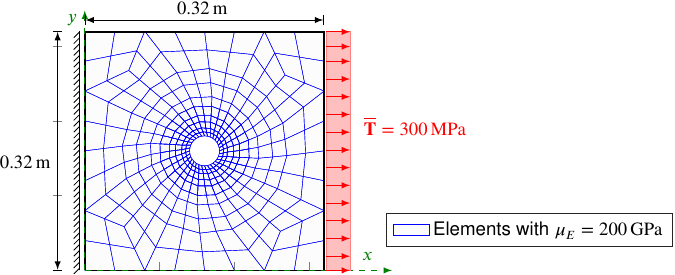}
	\caption{\textbf{Plate with a Hole:} The Geometry, boundary conditions, loading, and FE mesh. The blue mesh represents the FE mesh.}
	\label{fig:FE_Mesh_and_Geometry_and_Load}
\end{figure}

\noindent \textbf{Prior model:} The prior stochastic displacement is obtained using a linear elastic (LE) material model with a homogeneous Young's modulus $\meanFunc[E]{\spatialPoints} = \meanOf{E} = 200 \,\si{\GPa}$ and a standard deviation $\stdFunc[E]{\spatialPoints} = \stdOf{E} = 30 \,\si{\GPa}$. The Poisson's ratio is $\nu = 0.5$, making the material incompressible, meaning the volume does not change under loading. The displacements are calculated under the plane stress assumption, where the stress perpendicular to the plane is negligible, simplifying the problem to two dimensions. The thickness  is $0.01 \, \si{m}$, the radius of the hole is $0.02 \, \si{m}$, and the height and length are $0.32 \, \si{m}$. The geometry, the loading, and the FEM mesh are illustrated in \autoref{fig:FE_Mesh_and_Geometry_and_Load}.

\begin{figure}[!htb]
	\centering
	\subfloat[]{
		\centering
		\includegraphics{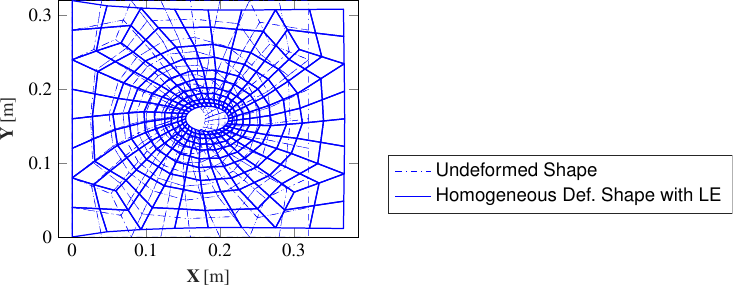}
		\label{subfig:plateWithHole_homogeneous_deformed}}
	\\
	\subfloat[$\meanVec{\stochDisp_{\text{f}}^x}$]{
		\centering
		\includegraphics{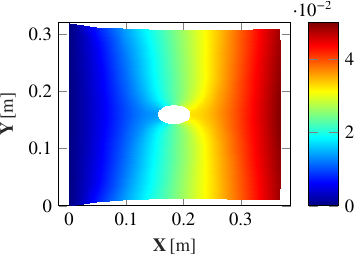}
		\label{subfig:plateWithHole_mean_dispX_contour}}
	\hspace*{1.0cm}
	\subfloat[$\stdVec{\stochDisp_{\text{f}}^x}$]{
		\centering
		\includegraphics{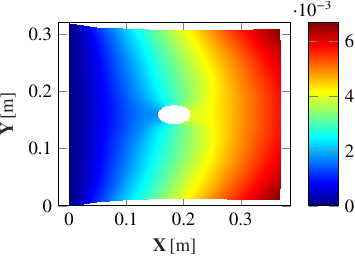}
		\label{subfig:plateWithHole_std_dispX_contour}}
	\caption{\textbf{Plate with a hole:} (a) The dashed blue lines represent the undeformed shape of the plate, and the blue lines represent the mean deformed shape with homogeneous Young's Modulus and LE material model. (b) Contour plot of prior (forecasted) mean displacement in $x$-direction $\meanVec{\stochDisp_{\text{f}}^x}$ with homogeneous Young's modulus and LE model. (c) Contour plot of prior (forecasted) standard deviation of displacement in $x$-direction $\stdVec{\stochDisp_{\text{f}}^x}$ with homogeneous Young's modulus and LE model.}
	\label{fig:plateWithHole_homogeneous_deformed_mean_dispX_std_dispX_contour}
\end{figure}

Note that the uncertain Young's modulus is log-normally distributed and considered as a weakly stationary random field. The parameters $\meanOf{\kappa} = 12.19$ and $\stdOf{\kappa} = 0.149$ of the weakly stationary Gaussian random field $\gaussRFs$ are obtained using \eqref{eq:mean_and_std_of_kappa_based_on_E}. In this detailed setup, we expand the Gaussian random field $\gaussRFs$ using the Mat\'ern $5/2$ kernel with a correlation length $\anIsotropicTwoDcorlength{\kappa}{x} = \anIsotropicTwoDcorlength{\kappa}{y} = 0.32/2$. Therefore, the number of KL terms, derived from \eqref{eq:explained_variance},  is determined to be $\nKlRvOfYoung = 13$. Considering equation \eqref{eq:POrder} with a univariate polynomial order $\sclp = 2$, the number of PC coefficients of the displacement is calculated to be $\pcOrderDisp = 105$.  The PC coefficients of the stochastic displacement $\stochDisp_{h}^x(\setYoungRVs) \in \bbR^{368 \times 1}$ and $\stochDisp_{h}^y(\setYoungRVs) \in \bbR^{368 \times 1}$ are then identified using \eqref{eq:PC_Coof_identify}. The mean deformed shape with LE material model and homogeneous Young's modulus, as well as the undeformed shape, are shown in \autoref{subfig:plateWithHole_homogeneous_deformed}. The contour plot of the mean displacement in $x$ direction $\meanVec{\stochDisp_{\text{a}}^x}$ is depicted in \autoref{subfig:plateWithHole_mean_dispX_contour} and the maximum of $\meanVec{\stochDisp_{\text{a}}^x}$ is $4.9825 \, \si{cm}$. The contour plot of standard deviation of displacement in $x$ direction $\stdVec{\stochDisp_{\text{a}}^x}$ is illustrated in \autoref{subfig:plateWithHole_std_dispX_contour} and the maximum of $\stdVec{\stochDisp_{\text{a}}^x}$ is $0.6676 \, \si{cm}$.\\

\begin{figure}[!htb]
	\centering
	\subfloat[]{
		\centering
		\includegraphics{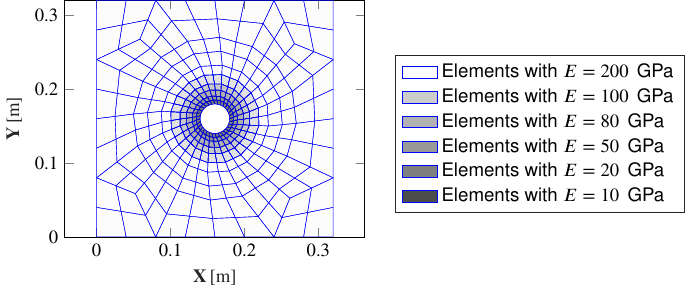}
		\label{subfig:plateWithHole_Mesh_nSen112}}
	\\
	\subfloat[]{
		\centering
		\includegraphics{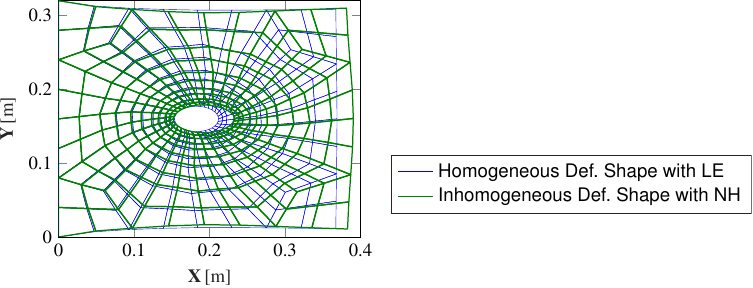}
		\label{subfig:plateWithHole_homogeneous_inhomogeneous_deformed}}
	\caption{\textbf{Plate with a Hole:} (a) The blue lines represent the FE mesh. (b) The blue lines represent the deformed shape with homogeneous Young's Modulus and Linear Elastic model. The green line represents the deformed shape with inhomogeneous Young's Modulus and Neo-Hooken model.}
	\label{fig:plateWithHole_homogeneous_inhomogeneous_deformed_shape}
\end{figure}

\noindent \textbf{Data generation:} To generate synthetic observation data, sensors were placed at various locations, particularly along the boundary. These sensor locations are depicted as black dots in \autoref{subfig:plateWithHole_Mesh_nSen112}. The synthetic observation data at these sensor locations were generated under two additional assumptions.
\begin{itemize}[noitemsep]
	\item \textbf{First}, we assume the material is rubber-like, leading us to select a hyperelastic material model for generating the experimental data, as opposed to the linear elastic assumption in the prior model.
	\item \textbf{Second}, we assume that drilling a hole in the plate has caused localized damage around the hole.
\end{itemize}
Regarding the first assumption, each hyperelastic model corresponds to a specific stored strain energy function. For incompressible isotropic materials, the stored strain energy function can be defined with three invariants $\sclI_1$, $\sclI_2$ and $\sclI_3$ of the right Cauchy-Green deformations tensor $\tenC$ as
\begin{equation}
	W(\tenC) = W(\sclI_1, \sclI_2, \sclI_3).
	\label{eq:strain_energy_function}
\end{equation}
The chosen hyperelastic model for generating the observation data is an incompressible \textit{Neo-Hookean} (NH) model, where the strain energy function is defined as
\begin{equation}
	W_{\text{NH}} = A_{10}(\sclI_1 - 3).
	\label{eq:strain_energy_function_NH}
\end{equation}
Here, the nonzero parameter $A_{10}$ is related to the shear modulus by $A_{10} = \sclG/2$. Given the Young's modulus $E$ and Poisson's ration $\nu = 0.5$, the shear modulus is $\sclG = E/3$, leading to $A_{10} = E/6$.
For the second assumption, the damage is represented by a reduction in Young's modulus in the vicinity of the hole. Specifically, Young's modulus in this region is not $200 \, \si{GPa}$ but is divided into five concentric rings, each with its own weakened Young's modulus. The innermost ring has the weakest Young's modulus, and as we move outward, the damage decreases, causing the Young's modulus to approach $200 \, \si{GPa}$. The elements for each ring are illustrated in \autoref{subfig:plateWithHole_Mesh_nSen112}. The deformed shape based on the inhomogeneous Young's modulus and NH material model is shown in  \autoref{subfig:plateWithHole_homogeneous_inhomogeneous_deformed}.

\begin{figure}[!htb]
	\vspace*{-0.5cm}
	\centering
	\subfloat[]{
		\centering
		\includegraphics{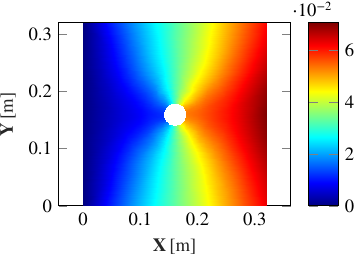}
		\label{subfig:plateWithHole_inhomogeneous_dispX_contour}}
	\hspace*{1.0cm}
	\subfloat[$\nsen = 11$]{
		\centering
		\includegraphics{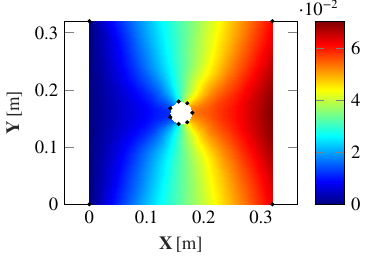}
		\label{subfig:plateWithHole_observation_dispX_contour_nSen11}}\\
	\subfloat[$\nsen = 32$]{
		\centering
		\includegraphics{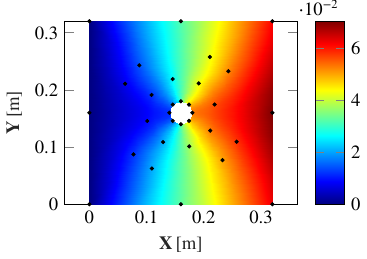}
		\label{subfig:plateWithHole_observation_dispX_contour_nSen32}}
	\hspace*{1.0cm}
	\subfloat[$\nsen = 112$]{
		\centering
		\includegraphics{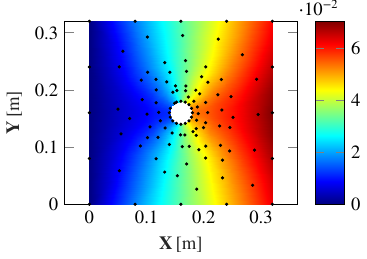}
		\label{subfig:plateWithHole_observation_dispX_contour_nSen112}}
	\caption{\textbf{Plate with a Hole:} (a) Contour plot of displacement in the $x$-direction for a plate with inhomogeneous Young's modulus and NH model $\vecu^{x}_{\text{NH}}$. \Rev{(b), (c), and (d) show the contour plot in the $x$-direction and sensor locations for different numbers of sensors $\nsen = 11$, $\nsen = 32$, and $\nsen = 112$. The black dots are the sensor locations.}}
	\label{fig:homogeneous_inhomogeneous_dispX_contour}
\end{figure}

Finally, the synthetic observation data $\SingleObs \in \mathbb{R}^{\nsen \times 2}$ \Rev{are generated for three cases with $\nsen = 11$, $\nsen = 32$, and $\nsen = 112$ sensor locations, each with $\nrep = 10$. The sensor locations for these cases are depicted in \autoref{subfig:plateWithHole_observation_dispX_contour_nSen11}, \autoref{subfig:plateWithHole_observation_dispX_contour_nSen32}, and \autoref{subfig:plateWithHole_observation_dispX_contour_nSen112}, respectively. The observation data are derived from the } NH model with inhomogeneous Young's modulus $\vecu_{\text{NH}} \in \mathbb{R}^{736}$ and contaminated with measurement noise, where the standard deviation is defined for each sensor as $\noisPcCof{1}, \dots, \noisPcCof{\nsen} = 0.001$. The contour plot of $\vecu_{\text{NH}}^x$ is depicted in \autoref{subfig:plateWithHole_inhomogeneous_dispX_contour}, where the maximum value of $\vecu_{\text{NH}}^x$ is $7.0400 \, \si{cm}$.\\

\noindent\textbf{Hyperparameter identification:} Given the observational data, the hyperparameters of the model-reality mismatch must be identified. For the model-reality mismatch, the Mat\'ern $5/2$ kernel is chosen, with correlation lengths of $\anIsotropicTwoDcorlength{d}{x} = \anIsotropicTwoDcorlength{d}{y} = 0.32/4$. \Rev{The number of KL terms $\nKlRvOfMrm$ in each direction is determined based on  \eqref{eq:explained_variance}. Specifically, if $\nsen = 11$, the number of KL terms is $\nKlRvOfMrm = 11$; for $\nsen = 32$, the number of KL terms is $\nKlRvOfMrm = 22$; and for $\nsen = 112$, we have $\nKlRvOfMrm = 32$.}

The unknown parameters for each KL mode, $\mrmKlCof{1}, \, \dots, \, \mrmKlCof{\nKlRvOfMrm}$, along with the scaling coefficients $\rhoCof^v$ in all directions, collectively referred to as the hyperparameters (as defined in \eqref{eq:hyperparameter_vector}), are identified by maximizing the marginal likelihoods according to \eqref{eq:maximizing_marginl_likelihood}. \Rev{The results of the identified hyperparameters for $\nsen = 11$ and $\nsen = 32$ are provided in \autoref{Appendix:Identified_Hyperparameters_for_Example_3} . Here, we present only the identified hyperparameters for $\nsen = 112$. \autoref{tab:identified_hyperparameters_2D_Md_32} shows} six identified hyperparameters out of a total of 32 for a model-reality mismatch with correlation lengths of $\anIsotropicTwoDcorlength{d}{x} = \anIsotropicTwoDcorlength{d}{y} = 8 \,\si{\cm}$ in a two-dimensional scenario. The values are given for two different dimensions, $v=1$ and $v=2$, with the following interpretations:

\begin{table}[H]
	\centering
	\begin{tabular}{ccccccc}
		\toprule
		Dimension & $\rhoCof^v $ & $ \mrmKlCof{1} $ & $ \mrmKlCof{2} $ & $ \mrmKlCof{3} $ & $ \mrmKlCof{4} $ & $ \mrmKlCof{5} $ \\\midrule
		$ v= 1 $  & $ 1.7969 $   & $ 0.01961 $      & $ 0.01643 $      & $ 0.01265 $      & $ 0.00953 $      & $ 0.00951 $      \\
		$ v= 2 $  & $ 1.1386 $   & $ 0.00245 $      & $ 0.00191 $      & $ 0.00188 $      & $ 0.00164 $      & $ 0.00164 $      \\
		\bottomrule
	\end{tabular}
	\caption{Identified $6$ of $33$ hyperparameters with $\anIsotropicTwoDcorlength{d}{x} = \anIsotropicTwoDcorlength{d}{y} =  8 \,\si{\cm}$ \Rev{for $\nsen = 112$} \label{tab:identified_hyperparameters_2D_Md_32}}
\end{table}

\begin{itemize}
	\item The larger scaling factor is $\rhoCof^v$ suggests that the model-reality mismatch is relatively larger in dimension $v=1$ compared to $v=2$.
	\item The KL coefficients decrease as the mode number increases, suggesting that higher-order modes contribute less to the mismatch uncertainty, which is typical in such expansions.
	\item The values for $v=1$ are consistently higher than for $v=2$, as the traction force is applied in the $x$-direction, making the displacement in the $x$-direction more dominant.
\end{itemize}
Based on the identified hyperparameters, we obtain the covariance matrix of the model-reality mismatch, $\covMat{\stochMrm}(\setMrmKlCofs)$\Rev{, for the three cases where $\nsen = 11$, $\nsen = 32$, and $\nsen = 112$. Using this, we compute} the K{\'a}lm{\'a}n gain based on \eqref{eq:kalman_gain} for each direction. The PC coefficients of displacement are then updated using the GMKF as described in \eqref{eq:PC_coefficiants_GMKF}. Once the PC coefficients are identified, the mean value and standard deviations of the posterior displacement can be obtained from \eqref{eq:Posterior_Mean_and_Cov_Disp_PC_Expansion}.

\begin{figure}[!]
	\vspace*{-1.5cm}
	\centering
	\subfloat[$\nsen = 11$]{
		\centering
		\includegraphics{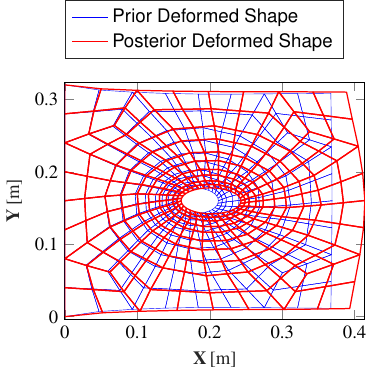}
		\label{subfig:plateWithHole_prior_posterior_deformed_shape_nSen11}}
	\hspace*{1.0cm}
	\subfloat[$\nsen = 11$]{
		\centering
		\includegraphics{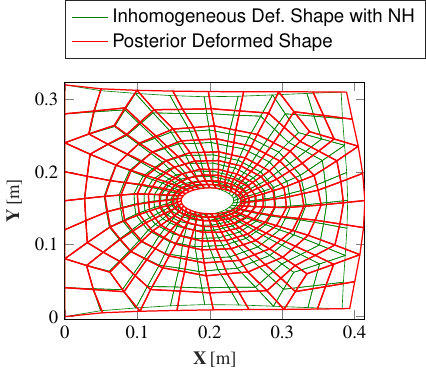}
		\label{subfig:plateWithHole_deformed_Y_nSen11}}\\
	\hspace*{-1cm}
	\subfloat[$\nsen = 32$]{
		\centering
		\includegraphics{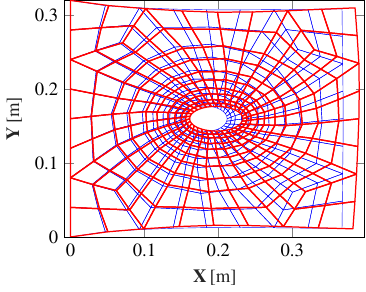}
		\label{subfig:plateWithHole_prior_posterior_deformed_shape_nSen32}}
	\hspace*{1.0cm}
	\subfloat[$\nsen = 32$]{
		\centering
		\includegraphics{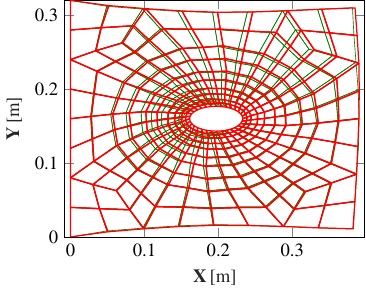}
		\label{subfig:plateWithHole_deformed_Y_nSen32}} \\
	\hspace*{-1cm}
	\subfloat[$\nsen = 112$]{
		\centering
		\includegraphics{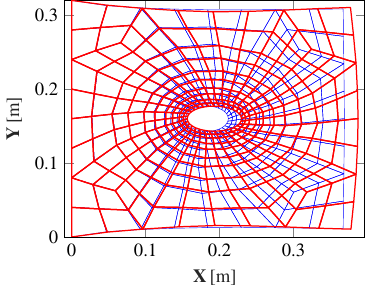}
		\label{subfig:plateWithHole_prior_posterior_deformed_shape_nSen112}}
	\hspace*{1.0cm}
	\subfloat[$\nsen = 112$]{
		\centering
		\includegraphics{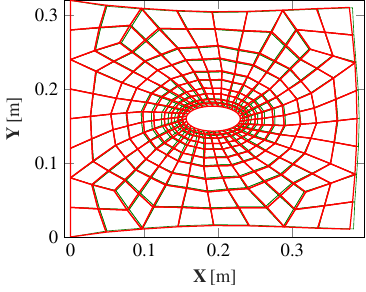}
		\label{subfig:plateWithHole_deformed_Y_nSen112}}
	\caption{\textbf{Plate with a Hole:} (a)\Rev{, (c) and (e)} The blue lines represent the prior mean deformed shape with homogeneous Young's Modulus and linear elastic material model. The red lines represent the posterior mean deformed shape \Rev{with $\nsen = 11$, $32$ and $112$}. (b)\Rev{, (d) and (f)} The green lines represent the deformed shape from the inhomogeneous Young's modulus and Neo-Hooke model. The red lines represent the posterior deformed shape \Rev{with $\nsen = 11$, $32$, and $112$}.}
	\label{fig:plateWithHole_prior_posterior_deformed_shape_Y}
\end{figure}

The mean prior deformed shape from \eqref{eq:Mean_and_Cov_Disp_PC_Expansion} and the mean posterior deformed shape \Rev{for the cases $\nsen = 11$, $32$ and $112$ are shown in \autoref{subfig:plateWithHole_prior_posterior_deformed_shape_nSen11}, \autoref{subfig:plateWithHole_prior_posterior_deformed_shape_nSen32} and \autoref{subfig:plateWithHole_prior_posterior_deformed_shape_nSen112}, respectively.} \Rev{As seen, the posterior displacement undergoes notable changes as the number of sensors increases. }  Moreover,  the mean posterior deformed shape differs noticeably from the prior, with larger deformations, particularly in the vicinity of the hole where Young's modulus is weaker than in the prior assumption. The posterior mean deformed shape can also be compared with the deterministic model, considered as a true response, which assumes a given inhomogeneous Young's modulus and a Neo-Hookean material model, as can be seen in \Rev{\autoref{subfig:plateWithHole_deformed_Y_nSen11}, \autoref{subfig:plateWithHole_deformed_Y_nSen32} and \autoref{subfig:plateWithHole_deformed_Y_nSen112} for the cases $\nsen = 11$, $32$ and $112$, respectively. It is evident that the posterior displacement fields obtained with $\nsen = 32$ and $\nsen = 112$ show a significantly better overlap with the true displacement compared to the case with $\nsen = 11$. This is because the sensors are placed uniformly overall on the domain. Moreover, it can be seen that the posterior displacement for $\nsen = 32$ and $\nsen = 112$ are more similar to the true response than the case with $\nsen = 11$.} The comparison \Rev{between the posterior deformed shape and true deformed shape} demonstrates the ability of the model-reality mismatch term to account for the two mismatches mentioned earlier: the assumption of a rubber-like material, which led to the selection of the NH model, and the inhomogeneous Young's modulus near the hole. Both of these factors were absent in the prior displacement provided within the statFEM framework but are effectively captured by the posterior displacement. 

\begin{figure}[!ht]
	\vspace*{-0.5cm}
	\centering
	\subfloat[$\nsen = 11$]{
		\centering
		\includegraphics{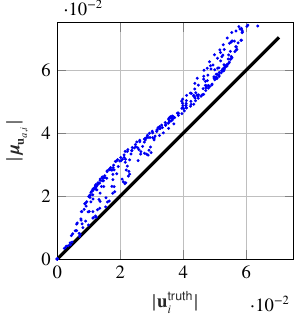}
		\label{subfig:plateWithHole_norm_nSen11}}
	\subfloat[$\nsen = 32$]{
		\centering
		\includegraphics{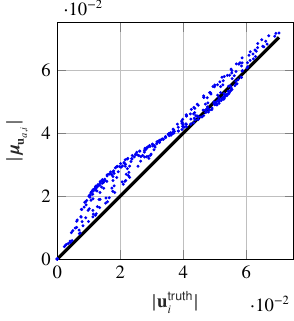}
		\label{subfig:plateWithHole_norm_nSen32}}
	\subfloat[$\nsen = 112$]{
		\centering
		\includegraphics{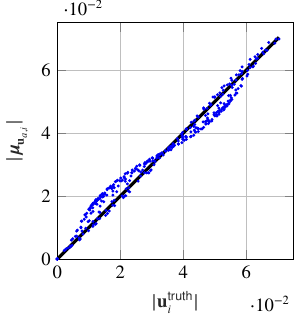}
		\label{subfig:plateWithHole_norm_nSen112}}
	\caption{\Rev{The comparison involves evaluating the norm of the displacement at every node between the true displacement field $\vecu_i^{\text{truth}}$, synthetically generated using the NH material model with an inhomogeneous Young's modulus around the hole, and the posterior displacement obtained from statFEM $\vecmu_{\vecu_{a,i}}$ with $i=1,\cdots,\nnode$ for different number of sensors.}}
	\label{fig:plateWithHole_norm}
\end{figure}

To better compare the true response and the posterior displacement, the Euclidean norm of the displacement at all nodes is \Rev{computed for the three cases where $\nsen = 11$, $\nsen = 32$, and $\nsen = 112$, and the results are} shown in \autoref{fig:plateWithHole_norm}. As illustrated, the displacement norms for all cases closely overlap with the true response\Rev{, with increasing agreement as the number of sensors increases. Notably, as $\nsen$ increases, the interpolated line approaches a slope of one, which indicates a stronger agreement between the true and posterior displacements.}

\begin{figure}[!htb]
	\vspace*{-0.5cm}
	\centering
	\subfloat[$\nsen = 11$ for $x$-direction]{
		\centering
		\includegraphics{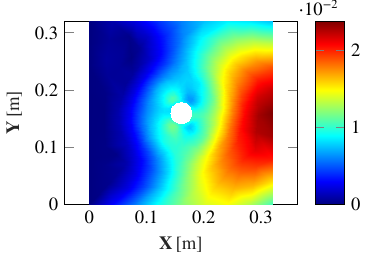}
		\label{subfig:plateWithHole_dispError_x_contour_nSen_11}}
	\hspace*{1.0cm}
	\subfloat[$\nsen = 11$ for $y$-direction]{
		\centering
		\includegraphics{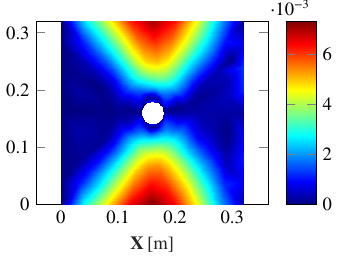}
		\label{subfig:plateWithHole_dispError_y_contour_nSen_11}}\\
	\subfloat[$\nsen = 32$ for $x$-direction]{
		\centering
		\includegraphics{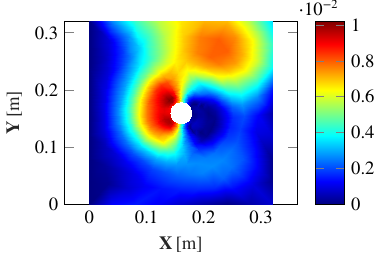}
		\label{subfig:plateWithHole_dispError_x_contour_nSen_32}}
	\hspace*{1.0cm}
	\subfloat[$\nsen = 32$ for $y$-direction]{
		\centering
		\includegraphics{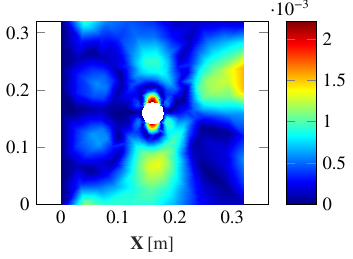}
		\label{subfig:plateWithHole_dispError_y_contour_nSen_32}}\\
	\subfloat[$\nsen = 112$ for $x$-direction]{
		\centering
		\includegraphics{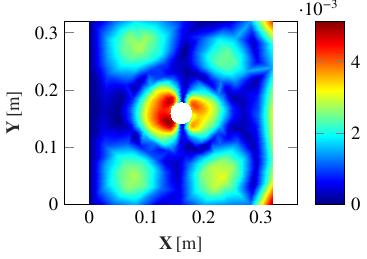}
		\label{subfig:plateWithHole_dispError_x_contour_nSen_112}}
	\hspace*{1.0cm}
	\subfloat[$\nsen = 112$ for $y$-direction]{
		\centering
		\includegraphics{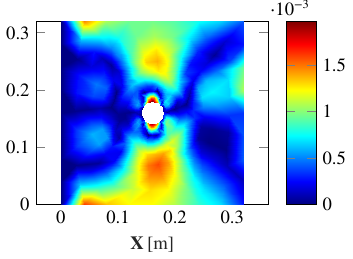}
		\label{subfig:plateWithHole_dispError_y_contour_nSen_112}}
	\caption{\textbf{Plate with a Hole:} (a)\Rev{, (c) and (e)} The absolute displacement error \Rev{for $\nsen = 11$, $32$, and $112$} in $x$-direction. (b)\Rev{, (d) and (f)} The absolute displacement error \Rev{for $\nsen = 11$, $32$, and $112$} in $y$-direction.}
	\label{fig:plateWithHole_dispError_y_contour_nSen}
\end{figure}

The absolute displacement error in the $x$ and $y$ directions \Rev{for the three cases are shown in \autoref{fig:plateWithHole_dispError_y_contour_nSen}. By comparing the results in \autoref{subfig:plateWithHole_dispError_x_contour_nSen_11}, \autoref{subfig:plateWithHole_dispError_x_contour_nSen_32}, and \autoref{subfig:plateWithHole_dispError_x_contour_nSen_112}, which depict the absolute displacement error in the $x$-direction, it is evident that the error decreases as the number of sensors increases. This demonstrates that a higher sensor density leads to a more accurate posterior displacement field in the $x$-direction. However, this error reduction is not as prominent in the $y$-direction, as seen in \autoref{subfig:plateWithHole_dispError_y_contour_nSen_11}, \autoref{subfig:plateWithHole_dispError_y_contour_nSen_32}, and \autoref{subfig:plateWithHole_dispError_y_contour_nSen_112}. The reason for this is that the applied force is primarily in the $x$-direction, meaning that displacement variations and their corrections are more significant along this axis. Consequently, the improvement in accuracy with increasing $\nsen$ is more noticeable in the $x$-direction than in the $y$-direction.}

\begin{figure}[!htb]
	\vspace*{-0.5cm}
	\centering
	\subfloat[$\meanVec{\stochDisp_{\text{a}}^x}$]{
		\centering
		\includegraphics{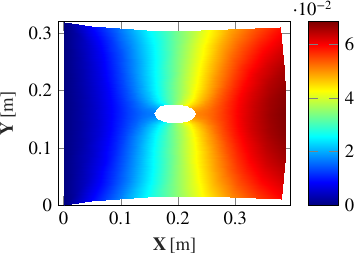}
		\label{subfig:plateWithHole_mean_dispX_Y_contour}}
	\hspace*{1.0cm}
	\subfloat[$\stdVec{\stochDisp_{\text{a}}^x}$]{
		\centering
		\includegraphics{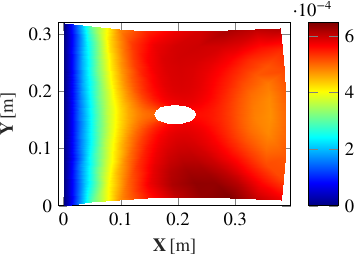}
		\label{subfig:plateWithHole_std_dispX_Y_contour}}
	\caption{\textbf{Plate with a Hole:} (a) Contour plot of posterior (assimilated) mean displacement in $x$-direction $\meanVec{\stochDisp_{\text{a}}^x}$ \Rev{with $\nsen = 112$}. (b) Contour plot of posterior (assimilated) standard deviation of displacement in $x$-direction $\stdVec{\stochDisp_{\text{a}}^x}$ \Rev{with $\nsen = 112$}.}
	\label{fig:plateWithHole_mean_dispX_Y_std_dispX_Y_contour}
\end{figure}

Finally, the mean and the standard deviation of the posterior displacement \Rev{for the case $\nsen = 112$} are shown in \autoref{subfig:plateWithHole_mean_dispX_Y_contour} and \autoref{subfig:plateWithHole_std_dispX_Y_contour}, respectively.\\

\noindent \Rev{\textbf{Discussion:} While the proposed statFEM framework provides a flexible and rigorous approach to incorporating model-reality mismatch, certain limitations and challenges remain. One key challenge is the identification of hyperparameters, which increases with the number of KL terms $\nKlRvOfMrm$ used to approximate the model error and the number of spatial dimensions. Moreover, as the number of sensors $\nsen$ increases, more KL terms $\nKlRvOfMrm$ are required, further adding to the complexity of the identification process. Additionally, the computational cost scales with both the number of sensors $\nsen$ and the number of sensor readings $\nrep$, as reflected in \eqref{eq:marginal_likelihood_for_multiple_observed_data} and a higher number of sensors also leads to a denser covariance matrix, which increases computational complexity. To mitigate this issue, low-rank approximations of dense covariance matrices, as proposed in \cite{duffin2022low}, offer a potential solution.}

\Rev{Another challenge is the selection of an appropriate kernel function for the model-reality mismatch. A smooth Gaussian kernel may fail to accurately capture the stochastic properties of the system in regions with high spatial variability, whereas an exponential kernel, although more localized, introduces a large number of hyperparameters and makes the identification process computationally expensive and potentially unstable. Thus, a careful selection of appropriate kernels is of utmost importance.}

\section{Conclusion}\label{sec:Conclusion_Outlook}
This paper introduced several critical improvements for the statFEM framework by introducing a sampling-free variant that removes the need for expensive online FEM computations. The proposed method handles the uncertainty in the prior by modeling the uncertain Young's modulus as a weakly stationary random field, which is expanded using the KL expansion. The resulting stochastic displacement field is described as a non-stationary, non-Gaussian random field that is efficiently approximated with PC expansions and evaluated by Smolyak sparse grids for dealing with high-dimensional integrals.

The mismatch between the model and reality is treated as a non-stationary Gaussian random field, \Rev{unlike previous works \cite{narouie2023inferring, girolami2021statistical}, where a stationary Gaussian model has been assumed. The model-reality mismatch is expanded with the KL basis, allowing for a more flexible and accurate representation of the spatially varying nature of discrepancies between the model and observed data.} The unknown parameters of the KL modes are treated as hyperparameters and are identified from observational data by maximizing the marginal likelihood. 

\Rev{Additionally, both the prior and posterior displacements are represented using PC expansion, which enables a consistent probabilistic formulation. The posterior displacement $\vecu^{a}$ is obtained by updating the PC coefficients of the prior displacement using the Gauss-Markov-K{\'a}lm{\'a}n filter. This ensures that the inferred displacement field incorporates spatially varying uncertainties in a statistically rigorous manner.}

Three benchmark examples validated the proposed method. First, the one-dimensional tension bar with a homogeneous Young's modulus showed the correctness of hyperparameter identification. The second example was also a one-dimensional tension bar but with the synthetic observational data collected from a model with heterogeneous Young's modulus, while the prior was assumed to be homogeneous \textemdash a typical engineering scenario. The results confirm that the model-reality mismatch can capture the difference in observed and prior displacement fields.

Lastly, the method was used on a two-dimensional plate with a hole. It was demonstrated that the misspecification term can capture the mismatch between the model and reality, especially close to the hole, where the displacement field is different because of the plate's drilling.

In summary, the introduced numerical examples show that the proposed sampling-free statFEM approach is highly flexible and computationally efficient, which makes it very suitable for practical applications like digital twins, where estimating mechanical states is very important. Using Polynomial Chaos and  Karhunen-Lo\`eve expansions, together with the Gauss-Markov-K{\'a}lm{\'a}n filter, allows analytically tractable posterior representations that are based on updated PC coefficients of displacements.

\section*{CRediT authorship contribution statement}
\textbf{Vahab Narouie (VN):} Conceptualization, Methodology, Formal Analysis, Software, Writing - Original Draft, Visualization. \textbf{Henning Wessels (HW):} Conceptualization, Writing - Review \& Editing, Funding Acquisition. \textbf{Fehmi Cirk (FC):} Conceptualization, Review \& Editing. \textbf{Ulrich R\"omer (UR):} Conceptualization, Methodology, Writing - Review \& Editing, Funding Acquisition.

\section*{Declaration of competing interest}
The authors declare that they have no known competing financial interests or personal relationships that could have appeared to influence the work reported in this paper.
\section*{Acknowledgement}\label{sec:Ackn}
The support of the German Research Foundation is gratefully acknowledged in the following projects:
\begin{itemize}
	\item DFG GRK2075-2 (VN, HW, UR): \textit{Modelling the constitutional evolution of building materials and structures with respect to aging}.
	\item DFG 501798687 (HW): \textit{Monitoring data-driven life cycle management with AR based on adaptive, AI-supported corrosion prediction for reinforced concrete structures under combined impacts}. Subproject of SPP 2388: \textit{Hundred plus - Extending the Lifetime of Complex Engineering Structures through Intelligent Digitalization}.
\end{itemize}
In both projects, methods are developed to link measurement data and physical models to improve structural health monitoring of aging materials and structures.
The authors would also like to thank Hermann Matthies for fruitful discussions on the topic.

\section*{Data availability}

\Rev{The source code supporting this study is available on GitHub and Zenodo \cite{narouie2025pc-based-statfem} to ensure reproducibility and facilitate further research.}

\begin{appendices}

	\section{Karhunen-Lo\`eve Expansion of Random Fields}\label{Appendix:Karhunen-Loeve_Expansion_of_Random_Fields}
	The KL expansion is a method for scalar-valued random fields based on a linear combination of orthogonal functions. The orthogonal functions are the eigenfunctions resulting from the spectral decomposition of the correlation function of the random field \cite{stefanou2009stochastic}. Note that Young's modulus is strictly positive and assumed to follow a lognormal distribution $\youngRFs = \exp{\gaussRFs}$ with the mean function $\meanFunc[E]{\spatialPoints} = \meanOf{E}$ and standard deviation function $\stdFunc[E]{\spatialPoints}= \stdOf{E}$. Here, $\gaussRFs$ is a weakly stationary Gaussian random field with $\meanFunc[\kappa]{\spatialPoints} = \meanOf{\kappa}$ and $\stdFunc[\kappa]{\spatialPoints} = \stdOf{\kappa}$ and a given infinitely-differentiable correlation function $\corFunc[\kappa]{\spatialPoints}$. The parameters $\meanOf{\kappa}$ and $\stdOf{\kappa}$ of the random field $\gaussRFs$ can be obtained as
	\begin{equation}
		\meanOf{\kappa} = \ln{\frac{\meanOf{E}^2}{\sqrt[]{\meanOf{E}^2+\stdOf{E}^2}}}, \quad  \stdOf{\kappa}^2 = \ln{1+\frac{\stdOf{E}^2}{\meanOf{E}^2}},
		\label{eq:mean_and_std_of_kappa_based_on_E}
	\end{equation}
	see \cite{ang2007probability,cao2017probabilistic}.
	The KL expansion of a second-order random field reads
	\begin{equation}
		\gaussRFs =  \meanFunc[\kappa]{\spatialPoints} + \stdOf{\kappa} \sum_{\scli=1}^{\infty} \sqrt{\eigenValue{i}} \, \eigenFunction[i]{\spatialPoints}\, \youngRVs{i} \approx \meanFunc[\kappa]{\spatialPoints} + \stdOf{\kappa} \sum_{\scli=1}^{\nKlRvOfYoung} \sqrt{\eigenValue{i}} \, \eigenFunction[i]{\spatialPoints} \, \youngRVs{i},
		\label{eq:kl_gaussian_random_variable_app}
	\end{equation}
	see \cite{karhunen1947ueber,loeve1948functions}, where $\eigenValue{i} \in [0,\infty), \, \eigenValue{1} \ge \eigenValue{2} \ge \cdots \ge 0$ are the non-increasing eigenvalues and $\eigenFunction{\scli} : \physicalDomain \rightarrow \bbR$ are the eigenfunctions of the correlation function $\corFunc[\kappa]{\spatialPoints}$. More precisely, the KL expansion expresses $\gaussRFs$ in terms of its mean $\meanFunc[\kappa]{\spatialPoints} = \expectOper[\gaussRFs]$ and a series expansion of orthogonal functions $\eigenFunction[i]{\spatialPoints}$ weighted by standard normal distributed random variables $\youngRVs{i}$.
	Following Mercer's theorem \cite{trees1968estimation}, the covariance function $\covFunc[\kappa]{\spatialPoints}$ has the following spectral decomposition:
	\begin{equation}
		\covFunc[\kappa]{\spatialPoints} = \sum_{\scli=1}^{\infty} \stdOf{\kappa}^2 \, \eigenValue{i} \, \eigenFunction[i]{\spatialPoints} \, \eigenFunction[i]{\spatialPoints'} \approx \sum_{\scli=1}^{\nKlRvOfYoung} \, \stdOf{\kappa}^2 \, \eigenValue{i} \, \eigenFunction[i]{\spatialPoints} \, \eigenFunction[i]{\spatialPoints'}.
		\label{eq:spectralDecKernel}
	\end{equation}
	Its eigenvalues $\eigenValue{i}$ and eigenfunctions $\eigenFunction[i]{\spatialPoints}$ are obtained from the solution of the homogeneous \textit{Fredholm integral equation} of the second kind given by
	\begin{equation}
		\int_{\physicalDomain} \corFunc[\kappa]{\spatialPoints} \eigenFunction[i]{\spatialPoints'} d \spatialPoints' = \eigenValue{i} \eigenFunction[i]{\spatialPoints}.
		\label{eq:secondKindFredholm}
	\end{equation}
	Correlation function $\corFunc[\kappa]{\spatialPoints}$ characterizes the specific correlation properties of a random field. A versatile type of correlation function used to model random fields is the Mat\'ern kernel, which is defined in \autoref{Appendix:Matern_kernels}.

	For practical implementation, the random field $\gaussRFs$ in \eqref{eq:kl_gaussian_random_variable} and its covariance function $\covFunc[\kappa]{\spatialPoints}$ in \eqref{eq:spectralDecKernel} are approximated after truncating the series at the $\nKlRvOfYoung$-th term, where $\nKlRvOfYoung \in \bbN^+$ is the number of the resulting independent standard Gaussian RVs. The truncation error and the explained variance are closely connected to the magnitude of the truncated eigenvalues, which in turn are significantly influenced by the correlation length of the field. Generally, a higher correlation among various locations within the spatial domain $\physicalDomain$ allows us to approximate the random field with smaller terms $\nKlRvOfYoung$. More precisely, the required number of terms can be determined based on the relation
	\begin{equation}
		\frac{\sum_{\scli=1}^{\nKlRvOfYoung} \eigenValue{i}}{\sum_{\scli=1}^{\infty} \eigenValue{i}} > 1-\epsilon,
		\label{eq:explained_variance}
	\end{equation}
	see \cite{wang2008karhunen,luthen2023spectral}, which expresses the explained variance, and a typical value for the threshold used in practice is $\epsilon = 0.001$.

	\section{Mat\'ern kernels}\label{Appendix:Matern_kernels}
	The general form of the Mat\'ern kernel based on the modified Bessel function is defined as follows \cite{matern1960spatial}:
	\begin{equation}
		\corFunc[\kappa, \scls]{\spatialPoints} = \frac{2^{1-\scls}}{\Gamma(\scls)} \Big(\frac{\sqrt{2\scls}\, \norm{\spatialPoints-\spatialPoints'}^2}{\scll_\kappa} \Big)^\scls \, \sclK_\scls \Big( \frac{\sqrt{2\scls}\, \norm{\spatialPoints-\spatialPoints'}^2}{\scll_\kappa}\Big),
		\label{eq:general_form_matern}
	\end{equation}
	where $\Gamma(\bullet)$ is the gamma function, $\sclK_\scls(\bullet)$ is the modified Bessel function of the second kind, $\scll_\kappa$ is the correlation length, and  $\scls$ is a parameter that controls the smoothness of the kernel. In this contribution, two kernels with different degrees of smoothness are considered. As $\scls \rightarrow \infty$, the Mat\'ern kernel approaches the well-known squared exponential kernel, and when $\scls = \frac{5}{2}$, the Mat\'ern $5/2$ kernel function can be expressed in the following simplified way.

	\begin{itemize}
		\item \textbf{1D:} For $\physicalDomain \subset \bbR$ the isotropic $1$-D squared exponential and the Mat\'ern $5/2$ kernels are expressed as follows:
		      \begin{equation}
			      \corFunc[\kappa, \scls \rightarrow \infty]{\spatialPoints} = \oneDsEcorFun{\kappa}{\sclX_\sclx},
			      \label{eq:one_diemensional_squared_exponential_covariance_function}
		      \end{equation}
		      \begin{equation}
			      \corFunc[\kappa, \scls = \frac{5}{2}]{\spatialPoints} = \oneDMaternCorFun{\kappa}{\sclX_\sclx},
			      \label{eq:one_diemensional_exponential_covariance_function}
		      \end{equation}
		      with $\spatialPoint = \sclX_\sclx$ and $\spatialPoint' = \sclX'_\sclx$.
		\item \textbf{2D:} For $\physicalDomain \subset \bbR^2$ the anisotropic $2$-D squared exponential and Mat\'ern $5/2$ kernels are given by
		      \begin{equation}
			      \corFunc[\kappa, \scls \rightarrow \infty]{\spatialPoints} = \anIsotropicTwoDseCorFun{\kappa}{\sclX},
			      \label{eq:two_diemensional_squared_exponential_covariance_function}
		      \end{equation}
		      \begin{equation}
			      \begin{split}
				      R_{\kappa, s= \frac{5}{2}}\left(\spatialPoints, \spatialPoints' \right) = & \left(1 + \frac{\sqrt{5}\norm{X_\sclx-X_\sclx'}}{l_{\kappa x}} + \frac{5\norm{X_\sclx-X_\sclx'}^2}{3l_{\kappa x}^2} \right)\left(1 + \frac{\sqrt{5}\norm{X_\scly-X_\scly'}}{l_{\kappa y}} + \frac{5\norm{X_\scly-X_\scly'}^2}{3l_{\kappa y}^2} \right) \\
				      & \exp{-\frac{\sqrt{5}\norm{X_\sclx-X_\sclx'}}{\anIsotropicTwoDcorlength{\kappa}{x}} -\frac{\sqrt{5}\norm{X_\scly-X_\scly'}}{\anIsotropicTwoDcorlength{\kappa}{y}} },
			      \end{split}
			      \label{eq:two_diemensional_exponential_covariance_function}
		      \end{equation}
		      where $\spatialPoints = (\sclX_\sclx, \sclX_\scly)$ and $\spatialPoints' = (\sclX'_\sclx, \sclX'_\scly)$ are the spatial position vectors of two points; $\anIsotropicTwoDcorlength{\kappa}{x}$ and $\anIsotropicTwoDcorlength{\kappa}{y}$ are the correlation lengths in $\direction{x}$ and $\direction{y}$ directions, respectively. In this contribution, only the isotropic correlation function is considered where $\anIsotropicTwoDcorlength{\kappa}{x} = \anIsotropicTwoDcorlength{\kappa}{y}= \isotropicTwoDcorlength{\kappa}$.
	\end{itemize}

	\section{Identified Hyperparameters for Example 1}\label{Appendix:Identified_Hyperparameters_for_Example_1}
	The identified hyperparameters for numerical example in \autoref{subsec:One-dimensional_Tension_Bar} for three other cases are shown in \autoref{tab:identified_hyperparameters_Md_2}, \autoref{tab:identified_hyperparameters_Md_3}, and \autoref{tab:identified_hyperparameters_Md_4}.
	\begin{enumerate}
		\item With $\oneDcorlength{\kappa} = \oneDcorlength{\stochMrm}= 100 \,\si{\mm}$, only two terms are required for the KL expansion of model-reality mismatch, i.e., $\nKlRvOfMrm = 2$. The identified hyperparameters are
		      \begin{table}[H]
			      \centering
			      \begin{tabular}{cccccc}
				      \toprule
				                       & Predefined & $\nrep = 1$  & $\nrep = 10$ & $\nrep = 100$ & $ \nrep = 1000$ \\\midrule
				      $ \rhoCof $      & $ 1.5 $    & $ 1.004977 $ & $ 1.984154 $ & $ 2.014405 $  & $ 2.015267 $    \\
				      $ \mrmKlCof{1} $ & $ 3.0 $    & $ 4.852275 $ & $ 4.162870 $ & $ 3.426323 $  & $ 3.103209 $    \\
				      $ \mrmKlCof{2} $ & $ 3.0 $    & $ 0.000007 $ & $ 4.000196 $ & $ 3.087127 $  & $ 2.998275 $    \\
				      \bottomrule
			      \end{tabular}
			      \caption{Identified hyperparameters with $ \isotropicTwoDcorlength{ \stochMrm } = 100 \,\si{\mm}$ \label{tab:identified_hyperparameters_Md_2}}
		      \end{table}

		\item Decreasing the correlation length to $\oneDcorlength{\kappa} = \oneDcorlength{\stochMrm}= 50 \,\si{\mm}$ increases the requirement to three KL terms, i.e. $\nKlRvOfMrm = 3$.
		      \begin{table}[H]
			      \centering
			      \begin{tabular}{cccccc}
				      \toprule
				                       & Predefined & $\nrep = 1$  & $\nrep = 10$ & $\nrep = 100$ & $ \nrep = 1000$ \\\midrule
				      $ \rhoCof $      & $ 1.5 $    & $ 1.363872 $ & $ 1.809648 $ & $ 1.817349 $  & $ 1.818990 $    \\
				      $ \mrmKlCof{1} $ & $ 3.0 $    & $ 5.218081 $ & $ 3.662685 $ & $ 3.482960 $  & $ 3.116315 $    \\
				      $ \mrmKlCof{2} $ & $ 3.0 $    & $ 0.947627 $ & $ 3.294290 $ & $ 2.978550 $  & $ 3.014204 $    \\
				      $ \mrmKlCof{3} $ & $ 2.5 $    & $ 0.000031 $ & $ 2.385881 $ & $ 2.541439 $  & $ 2.545575 $    \\
				      \bottomrule
			      \end{tabular}
			      \caption{Identified hyperparameters with $ \isotropicTwoDcorlength{ \stochMrm } = 50 \,\si{\mm}$ \label{tab:identified_hyperparameters_Md_3}}
		      \end{table}

		\item Further reduction to $\oneDcorlength{\kappa} = \oneDcorlength{\stochMrm}= 25 \,\si{\mm}$ needs four KL terms, i.e. $\nKlRvOfMrm = 4$.
		      \begin{table}[H]
			      \centering
			      \begin{tabular}{cccccc}
				      \toprule
				                       & Predefined & $\nrep = 1$  & $\nrep = 10$ & $\nrep = 100$ & $ \nrep = 1000$ \\\midrule
				      $ \rhoCof $      & $ 1.5 $    & $ 1.751046 $ & $ 1.731037 $ & $ 1.718646 $  & $ 1.720884 $    \\
				      $ \mrmKlCof{1} $ & $ 3.0 $    & $ 5.414911 $ & $ 3.797335 $ & $ 3.400352 $  & $ 3.042182 $    \\
				      $ \mrmKlCof{2} $ & $ 3.0 $    & $ 0.430689 $ & $ 3.338331 $ & $ 3.033997 $  & $ 2.980345 $    \\
				      $ \mrmKlCof{3} $ & $ 2.5 $    & $ 0.000129 $ & $ 2.454864 $ & $ 2.499716 $  & $ 2.517565 $    \\
				      $ \mrmKlCof{4} $ & $ 2.3 $    & $ 0.000033 $ & $ 2.135039 $ & $ 2.253934 $  & $ 2.338412 $    \\
				      \bottomrule
			      \end{tabular}
			      \caption{Identified hyperparameters with $ \isotropicTwoDcorlength{ \stochMrm } = 25 \,\si{\mm}$ \label{tab:identified_hyperparameters_Md_4}}
		      \end{table}
	\end{enumerate}

	\Rev{\section{Identified Hyperparameters for Example 3}\label{Appendix:Identified_Hyperparameters_for_Example_3}}
	\noindent\Rev{The identified hyperparameters for the numerical example in \autoref{subsec:two_dimensional_infinite_plate_with_a_hole_and_heterogeneous_young_s_modulus} for $\nsen = 11$ are shown in \autoref{tab:identified_hyperparameters_2D_Md_11} and for $\nsen = 32$ in \autoref{tab:identified_hyperparameters_2D_Md_22}.}
	\begin{table}[H]
		\centering
		\begin{tabular}{ccccccc}
			\toprule
			Dimension & $\rhoCof^v $ & $ \mrmKlCof{1} $ & $ \mrmKlCof{2} $ & $ \mrmKlCof{3} $ & $ \mrmKlCof{4} $ & $ \mrmKlCof{5} $ \\\midrule
			$ v= 1 $  & $ 1.2833 $   & $ 0.11968 $      & $ 0.09139 $      & $ 0.04587 $      & $ 0.02615 $      & $ 0.02380 $      \\
			$ v= 2 $  & $ 1.0089 $   & $ 0.01308 $      & $ 0.01076 $      & $ 0.00869 $      & $ 0.00630 $      & $ 0.00135 $      \\
			\bottomrule
		\end{tabular}
		\caption{\Rev{Identified $6$ of $12$ hyperparameters with $\anIsotropicTwoDcorlength{d}{x} = \anIsotropicTwoDcorlength{d}{y} =  8 \,\si{\cm}$ for $\nsen = 11$} \label{tab:identified_hyperparameters_2D_Md_11}}
	\end{table}

	\begin{table}[H]
		\centering
		\begin{tabular}{ccccccc}
			\toprule
			Dimension & $\rhoCof^v $ & $ \mrmKlCof{1} $ & $ \mrmKlCof{2} $ & $ \mrmKlCof{3} $ & $ \mrmKlCof{4} $ & $ \mrmKlCof{5} $ \\\midrule
			$ v= 1 $  & $ 1.6123 $   & $ 0.13763 $      & $ 0.13605 $      & $ 0.12462 $      & $ 0.11010 $      & $ 0.09985 $      \\
			$ v= 2 $  & $ 1.0226 $   & $ 0.01690 $      & $ 0.00432 $      & $ 0.00262 $      & $ 0.00165 $      & $ 0.00160 $      \\
			\bottomrule
		\end{tabular}
		\caption{\Rev{Identified $6$ of $24$ hyperparameters with $\anIsotropicTwoDcorlength{d}{x} = \anIsotropicTwoDcorlength{d}{y} =  8 \,\si{\cm}$ for $\nsen = 32$} \label{tab:identified_hyperparameters_2D_Md_22}}
	\end{table}

\end{appendices}
\bibliographystyle{unsrtnat}
\bibliography{literature}
\end{document}

%% file: packages/packages.tex
\usepackage{helvet} 
\usepackage{amsmath} 
\usepackage{amssymb} 
\usepackage[labelformat=simple]{subfig} 
\usepackage[numbers]{natbib} 
\usepackage{hyperref} 
\usepackage{siunitx} 
\usepackage{booktabs} 
\usepackage{float} 
\usepackage[page]{appendix} 
\usepackage{mathtools}
\usepackage{tikz} 
\usepackage{pgfplots} 
\pgfplotsset{compat=1.18}
\usepackage{mathrsfs} 
\usepackage{nomencl}
\usepackage{comment}
\usepackage{enumitem}
\usepackage{lineno}
\usepackage{placeins}

%% file: packages/definitions.tex
\newcommand{\Rev}{\textcolor{black}} 

\long\def\symbolfootnote[#1]#2{\begingroup \def\thefootnote{\fnsymbol{footnote}}\footnote[#1]{#2} \endgroup}

\newcommand{\scl}[1]{ \ensuremath{  #1  } }
\renewcommand{\vec}[1]{ \ensuremath { \mathbf{ #1 } } }
\newcommand{\gvec}[1]{ \ensuremath { \boldsymbol{ #1 } } }
\newcommand{\mat}[1]{ \ensuremath { \pmb{\mathbf{ #1 } } } }
\newcommand{\gmat}[1]{ \ensuremath {  \pmb{\mathbf{ #1 } } } }
\newcommand{\ten}[1]{ \ensuremath {  \pmb{\mathbf{ #1 } } }}

\newcommand{\scla}{\ensuremath{ \scl{a} }}
\newcommand{\sclb}{\ensuremath{ \scl{b} }}

\newcommand{\scld}{\ensuremath{ \scl{d} }}

\newcommand{\scli}{\ensuremath{ \scl{i} }}
\newcommand{\sclj}{\ensuremath{ \scl{j} }}
\newcommand{\sclk}{\ensuremath{ \scl{k} }}
\newcommand{\scll}{\ensuremath{ \scl{l} }}
\newcommand{\sclm}{\ensuremath{ \scl{m} }}
\newcommand{\scln}{\ensuremath{ \scl{n} }}

\newcommand{\sclp}{\ensuremath{ \scl{p} }}

\newcommand{\sclr}{\ensuremath{ \scl{r} }}
\newcommand{\scls}{\ensuremath{ \scl{s} }}

\newcommand{\sclw}{\ensuremath{ \scl{w} }}
\newcommand{\sclx}{\ensuremath{ \scl{x} }}
\newcommand{\scly}{\ensuremath{ \scl{y} }}

\newcommand{\sclA}{\ensuremath{ \scl{A} }}
\newcommand{\sclB}{\ensuremath{ \scl{B} }}

\newcommand{\sclF}{\ensuremath{ \scl{F} }}
\newcommand{\sclG}{\ensuremath{ \scl{G} }}

\newcommand{\sclI}{\ensuremath{ \scl{I} }}

\newcommand{\sclK}{\ensuremath{ \scl{K} }}
\newcommand{\sclL}{\ensuremath{ \scl{L} }}
\newcommand{\sclM}{\ensuremath{ \scl{M} }}

\newcommand{\sclP}{\ensuremath{ \scl{P} }}

\newcommand{\sclX}{\ensuremath{ \scl{X} }}

\newcommand{\veca}{\ensuremath{ \vec{a} }}
\newcommand{\vecb}{\ensuremath{ \vec{b} }}

\newcommand{\vecd}{\ensuremath{ \vec{d} }}
\newcommand{\vece}{\ensuremath{ \vec{e} }}

\newcommand{\vecg}{\ensuremath{ \vec{g} }}

\newcommand{\vecu}{\ensuremath{ \vec{u} }}

\newcommand{\vecw}{\ensuremath{ \vec{w} }}

\newcommand{\vecy}{\ensuremath{ \vec{y} }}
\newcommand{\vecz}{\ensuremath{ \vec{z} }}
\newcommand{\vecA}{\ensuremath{ \vec{A} }}
\newcommand{\vecB}{\ensuremath{ \vec{B} }}

\newcommand{\vecE}{\ensuremath{ \vec{E} }}

\newcommand{\vecL}{\ensuremath{ \vec{L} }}

\newcommand{\vecN}{\ensuremath{ \vec{N} }}
\newcommand{\vecO}{\ensuremath{ \vec{O} }}

\newcommand{\vecT}{\ensuremath{ \vec{T} }}

\newcommand{\vecX}{\ensuremath{ \vec{X} }}
\newcommand{\vecY}{\ensuremath{ \vec{Y} }}

\newcommand{\veczeta      }{\ensuremath{ \gvec{\zeta} }}

\newcommand{\veckappa     }{\ensuremath{ \gvec{\kappa} }}

\newcommand{\vecmu        }{\ensuremath{ \gvec{\mu} }}

\newcommand{\vecxi        }{\ensuremath{ \gvec{\xi} }}

\newcommand{\vecsigma     }{\ensuremath{ \gvec{\sigma} }}

\newcommand{\vecphi       }{\ensuremath{ \gvec{\phi} }}

\newcommand{\vecchi       }{\ensuremath{ \gvec{\chi} }}
\newcommand{\vecpsi       }{\ensuremath{ \gvec{\psi} }}

\newcommand{\vecPsi       }{\ensuremath{ \gvec{\Psi} }}

\newcommand{\vecvarrho    }{\ensuremath{ \gvec{\varrho} }}

\newcommand{\mata}{\ensuremath{ \mat{a} }}
\newcommand{\matb}{\ensuremath{ \mat{b} }}

\newcommand{\matd}{\ensuremath{ \mat{d} }}
\newcommand{\mate}{\ensuremath{ \mat{e} }}

\newcommand{\matu}{\ensuremath{ \mat{u} }}

\newcommand{\maty}{\ensuremath{ \mat{y} }}

\newcommand{\matA}{\ensuremath{ \mat{A} }}
\newcommand{\matB}{\ensuremath{ \mat{B} }}
\newcommand{\matC}{\ensuremath{ \mat{C} }}

\newcommand{\matH}{\ensuremath{ \mat{H} }}

\newcommand{\matK}{\ensuremath{ \mat{K} }}

\newcommand{\matY}{\ensuremath{ \mat{Y} }}

\newcommand{\matsigma     }{\ensuremath{ \pmb{\gmat{\sigma} }}}

\newcommand{\matphi       }{\ensuremath{ \gmat{\phi} }}

\newcommand{\matpsi       }{\ensuremath{ \gmat{\psi} }}

\newcommand{\matPsi       }{\ensuremath{ \gmat{\Psi} }}

\newcommand{\matvarrho    }{\ensuremath{ \gmat{\varrho} }}

\newcommand{\bbE}{\ensuremath{ \mathbb{E} }}

\newcommand{\bbK}{\ensuremath{ \mathbb{K} }}

\newcommand{\bbN}{\ensuremath{ \mathbb{N} }}

\newcommand{\bbP}{\ensuremath{ \mathbb{P} }}

\newcommand{\bbR}{\ensuremath{ \mathbb{R} }}

\newcommand{\bbU}{\ensuremath{ \mathbb{U} }}

\newcommand{\bbY}{\ensuremath{ \mathbb{Y} }}

\newcommand{\tenC}{\ensuremath{ \ten{C} }}

\newcommand{\tenP}{\ensuremath{ \ten{P} }}

\newcommand{\no}{\hspace{-0.14cm}\,}

\newcommand{\scaT}{\ensuremath{\mathrm{T}}}

\newcommand{\ex}{^{\xi}}

\newcommand{\ub}{_{\beta}}

\renewcommand{\d}[1]{\text{$\hspace{0.1cm}$d $\hspace{-0.11cm}#1$}}

\renewcommand{\sin}[1]{ \text{sin}\hspace{0.0cm}\left( {#1} \right) }
\renewcommand{\ln}[1]{\text{$\hspace{0.1cm}$ln$\left(#1\right)$}}
\renewcommand{\exp}[1]{\ensuremath{ \,\text{exp}{\left( #1 \right)} }}

\newcommand{\p}{\ensuremath{ ^{\prime} }}

\renewcommand{\det}[1]{\text{det$\left(#1\right)$}}

\newcommand{\s}{\ensuremath{ ^{2}  }}

\newcommand{\percent}{\ensuremath{ \%  }}

\renewcommand{\s}{\ensuremath{ ^*} }

\renewcommand{\t}{\ensuremath{ ^{\scaT} }}



%% file: packages/statFEM_definitions.tex
\makeatletter
\renewcommand{\fnum@figure}{\textbf{Fig. \thefigure}}
\makeatother

\DeclarePairedDelimiter{\norm}{\lvert}{\rvert}
\DeclareMathOperator{\diag}{diag}

\newcommand{\saSp}{\ensuremath{\Theta}}
\newcommand{\sigAl}{\ensuremath{\mathscr{F}}}
\newcommand{\prb}{\ensuremath{\bbP}}
\newcommand{\paraSpace}{\ensuremath{\bbK}}

\newcommand{\dobInner}[2]{\ensuremath{\left\langle #1, #2 \right\rangle}}
\newcommand{\physicalDomain}{\ensuremath{\Omega}}
\newcommand{\DimensionDomain}{\ensuremath{\scld}}
\newcommand{\spatialPoints}{\ensuremath{\vecX}}
\newcommand{\spatialPoint}{\ensuremath{\vecX}}
\newcommand{\outcomeSaSp}{\ensuremath{\theta}}
\newcommand{\orderFiniteDim}{\ensuremath{n}}
\NewDocumentCommand{\expectOper}{ O{} }{%
	\ifx&#1&%
	\ensuremath{\bbE}%
	\else
	\ensuremath{\bbE \left[#1\right]}%
	\fi
}

\NewDocumentCommand{\meanFunc}{ O{} m }{%
	\ifx&#1&%
	\ensuremath{\mu_{#2}}%
	\else
	\ensuremath{\mu_{#1}(#2)}%
	\fi
}

\newcommand{\meanOf}[1]{\ensuremath{\mu_{#1}} }
\NewDocumentCommand{\covFunc}{ O{} m }{%
	\ifx&#1&%
	\ensuremath{C_{#2}}%
	\else
	\ensuremath{C_{#1}(#2,#2')}%
	\fi
}

\newcommand{\covFuncD}[2]{\ensuremath{C_{#1}(#2)} }

\NewDocumentCommand{\corFunc}{ O{} m }{%
	\ifx&#1&%
	\ensuremath{R_{#2}}%
	\else
	\ensuremath{R_{#1}(#2,#2')}%
	\fi
}

\NewDocumentCommand{\stdFunc}{ O{} m }{%
	\ifx&#1&%
	\ensuremath{\sigma_{#2}}%
	\else
	\ensuremath{\sigma_{#1}(#2)}%
	\fi
}

\newcommand{\stdOf}[1]{\ensuremath{\sigma_{#1}} }
\newcommand{\suchAs}{\ensuremath{\,\,  | \,\,}}

\newcommand{\youngOutcomeS}{\ensuremath{\vecxi}}
\newcommand{\firstPK}{\tenP}
\newcommand{\stochDisp}{\ensuremath{\vecu}}
\newcommand{\preStochDisp}{\overline{\stochDisp}}
\newcommand{\preTranction}{\overline{\vecT}_{\vecN}}
\newcommand{\preDispOnBound}{\partial\physicalDomain_{\stochDisp}}
\newcommand{\preTracOnBound}{\partial\physicalDomain_{\firstPK}}

\newcommand{\unitNormal}{\vecN}
\newcommand{\setYoungRV}{ \ensuremath{ \vecxi(\outcomeSaSp)} }
\newcommand{\setYoungRVs}{ \ensuremath{ \vecxi } }

\newcommand{\youngRFs}{\ensuremath{E(\spatialPoints, \setYoungRVs)}}
\newcommand{\youngRV}[1]{ \ensuremath{ \xi_{#1}(\outcomeSaSp)} }
\newcommand{\youngRVs}[1]{ \ensuremath{ \xi_{#1}} }

\newcommand{\gaussRFs}{\ensuremath{\kappa(\spatialPoints, \setYoungRVs)}}
\newcommand{\eigenValue}[1]{\ensuremath{\lambda_{\scl#1}} }
\NewDocumentCommand{\eigenFunction}{ O{} m }{%
	\ifx&#1&%
	\ensuremath{\phi_{#2}}%
	\else
	\ensuremath{\phi_{#1}(#2)}%
	\fi
}
\newcommand{\eigenPairs}[1]{ \ensuremath{ \{\eigenValue{#1},\eigenFunction{#1}{(\spatialPoints)}\} } }
\newcommand{\gaussRVECs}{\ensuremath{\veckappa_{h}(\setYoungRVs)}}
\newcommand{\gaussRVARs}[1]{\ensuremath{\kappa(\spatialPoints_{#1}, \setYoungRVs)}}
\newcommand{\youngRVECs}{\ensuremath{\vecE_{h}(\setYoungRVs)}}

\newcommand{\oneDcorlength}[1]{\ensuremath{\scll_{#1}} }
\newcommand{\anIsotropicTwoDcorlength}[2]{\ensuremath{\scll_{#1\scl#2}} }
\newcommand{\oneDsEcorFun}[2]{\ensuremath{\exp{-\frac{\norm{#2-#2'}^2}{2\oneDcorlength{#1}^2}}} }

\newcommand{\oneDMaternCorFun}[2]{\ensuremath{\left(1+ \frac{\sqrt{5}\norm{#2-#2'}}{\oneDcorlength{#1}} + \frac{5\norm{#2-#2'}^2}{3\oneDcorlength{#1}^2} \right)\exp{-\frac{\sqrt{5}\norm{#2-#2'}}{\oneDcorlength{#1}}}} }

\newcommand{\isotropicTwoDcorlength}[1]{\ensuremath{\scll_{#1}} }
\newcommand{\anIsotropicTwoDseCorFun}[2]{\ensuremath{\exp{-\frac{\norm{#2_\sclx-#2_\sclx'}^2}{2\anIsotropicTwoDcorlength{\kappa}{x}^2} -\frac{\norm{#2_\scly-#2_\scly'}^2}{2\anIsotropicTwoDcorlength{\kappa}{y}^2} }} }

\newcommand{\direction}[1]{\ensuremath{\scl#1}}
\newcommand{\nKlRvOfYoung}{\ensuremath{\sclM_{\kappa}}}

\newcommand{\pcOrderDisp}{ \ensuremath{ \sclP_{\stochDisp} } }
\newcommand{\pcCof}[2]{\ensuremath{ #1_{#2}}}
\newcommand{\pcMultiHermite}[2]{\ensuremath{ \Psi_{#1}(#2)}}

\newcommand{\meanVec}[1]{\ensuremath{\vecmu_{#1}}}
\newcommand{\stdVec}[1]{\ensuremath{\vecsigma_{#1}} }
\newcommand{\covMat}[1]{\ensuremath{\matC_{#1}} }
\newcommand{\hermiteDobInner}[2]{\ensuremath{\left\langle \Psi_{#1}^2\big(#2\big) \right\rangle} }
\newcommand{\nnode}{\scln_{\text{nod}}}                           
\newcommand{\dispPcCofMatrix}{\ensuremath{\matu^v}}
\newcommand{\dispPcHerVec}{\ensuremath{\vecPsi}}
\newcommand{\nGL}{\scln_{\text{gl}}}
\newcommand{\wGL}{\sclw_{\scln}}

\newcommand{\SingleObs}{\ensuremath{\vecy_{ \sclr }}}

\newcommand{\surrSingleObs}{\ensuremath{\vecy_{\text{f}}}}
\newcommand{\gaussianPartSingleObs}{\ensuremath{\tilde{\vecy}}}
\newcommand{\multipleObs}{\ensuremath{\matY}}
\newcommand{\obsOperator}[1]{\ensuremath{\vecO}\big(#1\big)}
\newcommand{\stochMrm}{\ensuremath{\vecd}}
\newcommand{\stochNoise}{\ensuremath{\vece}}
\newcommand{\rhoCof}{\ensuremath{\rho } }
\newcommand{\Hproj}{ \ensuremath{\matH } }

\newcommand{\mrmKlCof}[1]{\ensuremath{\sigma_{\stochMrm^v,\scl#1} } }
\newcommand{\mrmKlCofVec}[1]{\ensuremath{\vecsigma_{\stochMrm^v,\scl#1} } }
\newcommand{\setMrmKlCof}[1]{\ensuremath{\vecsigma_{\stochMrm^{ #1 }} } }
\newcommand{\setMrmKlCofs}{\ensuremath{\vecsigma_{\stochMrm} } }
\newcommand{\setMrmRV}{ \ensuremath{ \vecchi(\outcomeSaSp)} }
\newcommand{\setMrmRVs}{ \ensuremath{ \vecchi} }
\newcommand{\nKlRvOfMrm}{\ensuremath{\sclM_{\stochMrm}}}
\newcommand{\noisRV}[1]{\ensuremath{ \zeta_{#1} } }
\newcommand{\setNoisRV}{\ensuremath{ \veczeta(\outcomeSaSp) } }
\newcommand{\setNoisRVs}{\ensuremath{ \veczeta } }
\newcommand{\veczero}{\ensuremath{ \vec{0} }}
\newcommand{\noisPcCof}[1]{\ensuremath{ \vecsigma_{\stochNoise^v, \scl#1}  } }
\newcommand{\noisPcCofT}[1]{\ensuremath{ \vecsigma_{\stochNoise^v, \scl#1}^\top  } }
\newcommand{\noisPcCofMat}{\ensuremath{ \matsigma_{\stochNoise^v} } }
\newcommand{\eigenVec}[1]{ \ensuremath{ \vecphi_{\scl#1}} }
\newcommand{\nsen}{\scln_{\text{sen}}}
\newcommand{\ngdof}{\scln_{\text{gdof}}}                            
\newcommand{\ngsen}{\scln_{\text{gsen}}}                            

\newcommand{\nrep}{\scln_{\text{r}}}                            
\newcommand{\eigenMat}{ \ensuremath{ \matphi } }
\newcommand{\mrmKlCofMatrix}{\ensuremath{\matsigma_{\stochMrm^v} } }
\newcommand{\exPcOrder}{ \ensuremath{ \sclP_{\alpha} } }
\newcommand{\exPcMultiHermite}[2]{\ensuremath{ \hat{\Psi}_{#1}\big(#2\big)}}
\newcommand{\exPcCof}[2]{\ensuremath{ \hat{#1}_{#2}}}

\newcommand{\exDispPcCofMatrix}{\ensuremath{\hat{\matu}}}
\newcommand{\exMrmPcCofMatrix}{\ensuremath{\hat{\matd}}}
\newcommand{\exNoisPcCofMatrix}{\ensuremath{\hat{\mate}}}
\newcommand{\exObsPcCofMatrix}{\ensuremath{\hat{\maty}}}
\newcommand{\exPcVec}{\ensuremath{\hat{\vecPsi}}}

\newcommand{\exHermiteDobInner}[4]{\ensuremath{\left\langle \hat{\Psi}_{#1}^2\big(#2, #3, #4 \big) \right\rangle} }
\newcommand{\nMC}{ \ensuremath{ \scln_{\text{sm}} }}
\newcommand{\PDF}[2]{\ensuremath{f_{#1} (#2)}}

\newcommand{\kgain}{\ensuremath{\matK}}
\newcommand{\trueDisp}{\ensuremath{\vecz}}

\newcommand{\lebesgueFunction}{\ensuremath{\vecL}}

\newcommand{\vecEmpty}{\ensuremath{\boldsymbol{\varnothing}} }
\newcommand{\mrmRV}[1]{\ensuremath{ \chi_{#1} } }

\newcommand{\hypVec}{\ensuremath{\vecw} }
\newcommand{\identifiedHypVec}{\ensuremath{\vecw^*} }
\newcommand{\mPsi}{\Psi}
\DeclareMathOperator*{\argmax}{arg\,max}                      
\DeclareMathOperator*{\argmin}{arg\,min}                      

